**Late Neandertals in Central Italy. High-resolution chronicles from Grotta dei Santi (Monte Argentario - Tuscany)**

Adriana Moroni [a, b, c], Giovanni Boschian [d,*], Jacopo Crezzini [a, c], Guido Montanari-Canini [e], Giulia Marciani [a, e, f], Giulia Capecchi [a, c], Simona Arrighi [g, a, c], Daniele Aureli [a, h, c], Claudio Berto [e], Margherita Freguglia [a], Astolfo Araujo [b,i], Sem Scaramucci [a, c], Jean Jacques Hublin [j,] Tobias Lauer [j], Stefano Benazzi [g], Fabio Parenti [b, k], Marzia Bonato [l], Stefano Ricci [a], Sahra Talamo [j], Aldo G. Segre [b], Francesco Boschin [a, c], Vincenzo Spagnolo [a, c].

a. Dipartimento di Scienze Fisiche, della Terra e dell'Ambiente – Unità di Ricerca di Preistoria e Antropologia – Università di Siena – Italy.
b. Istituto Italiano di Paleontologia Umana – Anagni – Italy.
c. Centro Studi sul Quaternario Onlus.
d. Dipartimento di Biologia – Università di Pisa – Italy.
e. Dipartimento di Studi Umanistici – Sezione di Scienze Preistoriche e Antropologiche – Università di Ferrara – Italy.
f. Departamento de Historia e Historia del Arte, Universitat Rovira I Virgili Tarragona - Spain.
g. Department of Cultural Heritage, University of Bologna, Via degli Ariani 1, 48121 Ravenna - Italy.
h. UMR 7041 ArScAn équipe AnTET.MAE,21, avenue de l'Université, 92023 Nanterre - France.
i. Laboratório Interdisciplinar de Pesquisas em Evolução, Cultura e Meio Ambiente, Museu de Arqueologia e Etnologia da Universidade de São Paulo – Brazil.
J. Max Planck Institute for Evolutionary Anthropology – Department of Human Evolution – Leipzig – Germany.
k. Universidade de Federal do Paraná, Curitiba - Brazil.
l. Civico Museo Archeologico di Camaiore – Italy.
* Corresponding author. E-mail address: gboschian@biologia.unipi.it

**Abstract:**

Most of the Middle Palaeolithic evidence of Central Italy still lacks a reliable chrono-cultural framework mainly due to research history. In this context Grotta dei Santi, a wide cave located on Monte Argentario, on the southern coast of Tuscany, is particularly relevant as it contains a very well preserved sequence including several Mousterian layers. Research carried out at this site in the last years (2007e2017) allowed for a preliminary estimation of its chronology based on a set of radiometric determinations which place the investigated sequence in the time interval between 50 and 40 ka BP. Alongside the chronological issue, this paper mainly focuses on the geoarchaeological and zooarchaeological (micro and macro fauna) studies carried out on the materials retrieved during the 2007e2014 excavation fieldworks. The results of these studies are consistent with those from the radiometric chronology. A state of art concerning the MIS3 Italian sites is also provided in order to highlight the key role Grotta dei Santi may play in the assessment of late Neandertals' behaviour within the framework of the Middle to Upper Palaeolithic transition of Central Italy.

**1. Introduction**

Crucial bio-cultural changes took place in western Eurasia be-tween 50 and 35 ka ago, eventually reaching their climax when the last resident Neandertal populations were replaced by the intrusive so-called "Anatomically Modern Humans" (AMHs) (Hublin, 2015), whomay have arrived in Europe as early as 45 ka ago or even before (Douka et al., 2014; Zanchetta et al., 2018). This model is supported by palaeontological and genetic evidence setting the earliest successful dispersal of AMHs from Africa into Eurasia around 60-50 ka ago, along the Levant corridor (Fu et al., 2014; Hershkovitz et al., 2015; Pagani et al., 2015; Posth et al., 2016). Although the Middle to Upper Palaeolithic transition has been a key research topic for the last twenty years, the intricate scenario emerging from the archaeological record still poses many unsolved questions, due also to the general scarcity of human remains. These questions especially concern the nature of the makers of transitional techno-complexes but also, their behaviour and relationships in terms of possible cohabitation and reciprocal cultural influences, as well as the exact timing of the substitution of Neandertals by AMHs, especially when the regional detail is considered (Higham et al., 2014; Hublin et al., 2012). Recent studies based on improved dating methods produced new fine-tuned chronologies resulting in a more reliable spatio-temporal framework for the Middle to Upper Palaeolithic transition. The demise of the Neandertal populations took place apparently within the same period (41,030e39,260 cal BP - at 95.4% probability) throughout Europe, although some regional variability is reported (Highamet al., 2014). When considering the Italian territory as a whole, this issue is particularly felt for Central Italy where stratified sequences referred to this period are scarce and have not been included in recent research projects. Different lithic traditions (Mousterian, Uluzzian, Proto-Aurignacian) were contemporarily present within the Italian territory in the time span between 45 and 39 ka ago. The Uluzzian and the Proto-Aurignacian spread into Northern Italy as early as 44.4e42.8 ka (Riparo del Broion and Grotta di Fumane in Veneto - Benazzi et al., 2015; Peresani et al., 2016, in press) and 42.7e41.6 ka cal BP respectively (Riparo Mochi in Liguria - Douka et al., 2012; Grimaldi and Santaniello, 2014), possibly when the latest Mousterians still survived in the same areas (Higham et al., 2014; Negrino and Riel-Salvatore, 2018). In Southern Italy, the final Mousterian possibly overlapped for about 3000 years with the Uluzzian e recently referred to AMHs (Benazzi et al., 2011; Moroni et al., in press) e although radiocarbon determinations obtained for the Mousterian in this region should be considered as minimum ages according to Douka et al. (2014).

Times and modes of the transition are more unclear in Central Italy, because reliable chronological and stratigraphic data are scarce. Collectively Middle Palaeolithic contexts are frequent throughout the area, but not homogeneously distributed. Moreover, it is often difficult to ascertain which ones belong to the final phase because no geo-chronometry data is available. Only one cave site e Grotta La Fabbrica in Tuscany (Villa et al., 2018) e includes a stratigraphic sequence spanning the whole transition period (Mousterian, Uluzzian, Proto-Aurignacian), whereas several Tuscan open-air sites yielded Mousterian, Uluzzian and Proto-Aurignacian artefacts in mixed surface contexts (Palma di Cesnola, 1993).

These points show that it is vital to collect e in Tuscany as well as all over Central Italy e new stratigraphic and chronologic information by modern systematic excavations and cutting-edge dating methods. Grotta dei Santi (Crezzini and Moroni, 2012; Freguglia et al., 2007; Moroni et al., 2015; Moroni Lanfredini et al., 2010; Spagnolo, 2017), a cave located on the southern coast of Tuscany, is expected to contribute to solving part of these issues, as it includes a largely undisturbed Middle Palaeolithic sequence yielding archaeological and paleoenvironmental data, that are almost unique in Central Italy. According to first radiocarbon dating reported here, this sequence ranges approximately from 50 to 40 ka cal BP. It is therefore an ideal site to gather reliable information about the cultural and spatio-temporal context of the last Neandertals, immediately before and during the Middle to Upper Palaeolithic transition. This paper presents preliminary data about the stratigraphy and radiometric chronology of the cave infill, and presents the paleo-environmental context (macro and micro fauna studies) of the Mousterian occupation of the cave (excavation seasons 2007e2016).

## 2. The site

### 2.1. Physical landscape

Grotta dei Santi (or Grotta di Cala dei Santi) is a wide cave located on the southern side of the Monte

Argentario promontory, on the Tyrrhenian Sea coast of Southern Tuscany (Central Italy). This promontory is a steep and almost isolated mount significantly higher (maximum height 615m a.s.l.) than the contiguous mainland area (Fig. 1). In fact, this mount is an island connected to the mainland by two thin strips of sand, which enclose the Orbetello lagoon. During the Pleistocene, Mt. Argentario was intermittently situated well within the mainland, depending on the eustatic low-and highstands of the sea-level (Fig. 1).

The geology of Mt. Argentario is characterised by a Late Triassic basement of phyllites covered by overthrusted limestones/dolomitic limestones and Cretaceous limestones. The structure is complicated by secondary overthrusting and intensive faulting and erosion, which put into evidence a complex outcrop pattern. Similar rock types crop out along the coast in the vicinity of Monte Argentario, whereas the Middle Pleistocene volcanic complexes of Mt. Amiata (trachydacites and subordinate olivin-latites/shoshonites) and Middle to Late Pleistocene Mts. Vulsini (KS -trachybasalt totrachyte and HKS-leucitite and leucite tephrite to phonolite) (Peccerillo, 2005) are situated more to the inland, to the NE and E respectively. Consequently, the rivers transport sediments rich in volcanic minerals that are subsequently discharged into the coastal plain facing Monte Argentario and into the nearby area of the Tyrrhenian Sea. Grotta dei Santi develops into the "Calcare cavernoso" formation, a Triassic dolomitic limestone. In the southern Argentario area, this formation is thickly and irregularly layered, and is characterised by a breccia-like facies with abundant clay and/or generally silicate impurities, as well as some allochthonous rock clasts. Interestingly, some wide patches of moderately pedogenised aeolian sand with few volcanic grains crop out on the mountain side above the cave, approximately from 100 to 200m a.s.l.

Caves are common in the limestone formations of the promontory (22 surveyed at present), but a half of these are now underwater or partly invaded by the sea. To date, Grotta dei Santi is the only one which yielded evidence of Palaeolithic remains. Grotta dei Santi is a water-mixing dissolution cavity located almost at sea-level (2m a.s.l.), at the foot of a limestone falaise about 50mhigh, in a coast area that can now be accessed only by boat (Fig. 2). The cave entrance is about 12mwide and opens inside a small cove resulting from the collapse of part of the cavity, which probably belonged to a much larger old underground system. Some nichese possibly parts of the same old system e that can be observed at different heights on the falaise, still include speleothem remains and finely layered sand-size clastic sediments. The entrance of the cave is marked by an irregular step, about 0e2m high, variously shaped in the limestone bedrock by sea erosion and karstic dissolution. The drip-line is up to 6m high and situated approximately above the rock step, overhanging the sea on the northern side of the entrance, and leaving the step uncovered on the southern side.

The shape of the cave (Fig. 2) is roughly rectangular, about 48 35mwide, with the major axis oriented NE-SW; the entrance is situated in the north-eastern corner. The ceiling is slightly domed and with dissolution pockets. Minor passages extend for some tens of metres from the bottom wall into the rock mass.

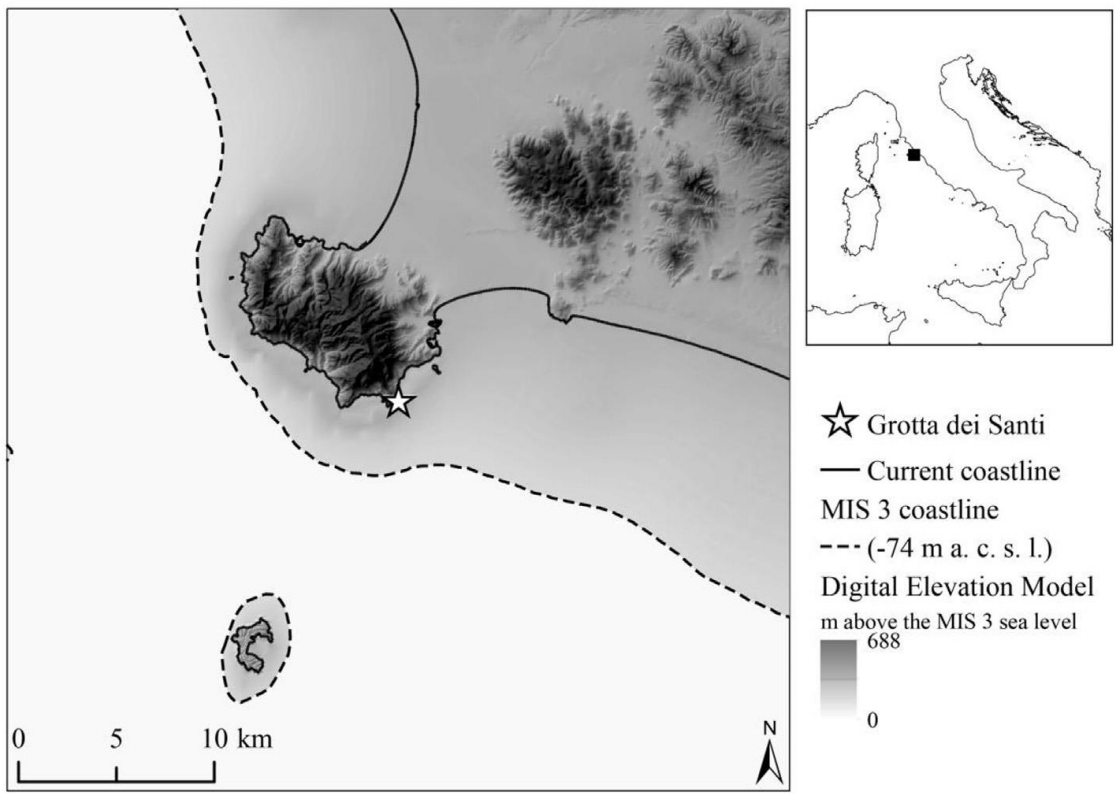

**Fig. 1.** Grotta dei Santi. Localization of the site on Mt. Argentario (Tuscany, Italy) and reconstruction of the possible coast-line during the MIS 3 cave occupation phase (Waelbroeck et al., 2002; Antonioli, 2012). This model was adjusted taking into account the local uplift rate (that here it results relatively modest) þ 0.2 mm/a in the last 50 ka (Ferranti et al., 2006; Spagnolo, 2017). Source of Digital Elevation Model: Regional Information System of Tuscany (http://www.regione.toscana.it/-/geoscopio). Source of Bathymetric Model: European Marine Observation and Data Network (EMODnet). GIS elaboration of geodata: Vincenzo Spagnolo.

The cave is partly filled by clastic and chemical sediments. In the southern (inner) part, these are shaped into a flat terrace at about 10m above the present-day sea-level. To the north, the terrace floor slopes gently up to about 15m a.s.l., where a group of large stalagmite bosses marks its top.

A subvertical scarp shaped bywave action cuts the infill down to the bedrock and marks the eastern limit of the terrace, facing the entrance of the cave. Large part of the scarp is a chaotic accumulation of very large boulders cropping out from the bottom of the infill and complexly interlayered by overgrown stalagmites and stalagmite crusts.

Two erosion channels (or "corridors") run along the northern and southern walls of the cave. The northern one is rather narrow (1e2m) and deep (2e5m), and partly filled by recent speleothems. The southern one is wider (3e6m) and shallower (2e3m) and exposes the sequence of fine clastic sediments that infill the cave, down to a major level of speleothem crusts and bosses. The archaeological excavations were located in this area, starting from the naturally exposed sequence, which can be used as a natural explorative sondage.

**2.2. Research history**

Worked flints had been recorded in the cave since the mid 19[th] century (Merciai, 1910; Nicolucci, 1869;

Salvagnoli and Marchetti, 1843); on June 16th, 1951 the site was visited by A.G. Segre of the Italian Institute of Human Palaeontology, who provided a synthetic stratigraphic description of the cave deposit (Segre, 1959). This occurrence of lithics included in a well-preserved geological sequence suggested that interdisciplinary research on this site may provide high quality information about the behavioural evolution of Neandertals and its chronology in Central Italy. Investigations started in 2007, in agreement with the Archaeological Office of Tuscany (Freguglia et al., 2007; Moroni Lanfredini et al., 2010) and have been conducted by the University of Siena in collaboration with the Italian Institute of Human Palaeontology and the University of Pisa. In 2013 the University of S~ao Paulo (Brazil) joined the research team of Grotta dei Santi and a collaboration with the Max Planck Institute for Evolutionary Anthropology of Leipzig was established in 2014 (Marciani et al., 2018).

To date several different units, both sterile and anthropogenic, were identified (20.3.3e105, 20.3.3e106, 20.3.3e107, 20.3.2e110, 20.3.2e111, 20.3.1e125, 20.3.1e150, 20.2e1004; see Figs. 4 and 5 and section 4.1). Only the upper anthropogenic units 20.3.2e110, 20.3.2e111, and 20.3.1e150 have been extensively investigated. These are generally characterised by recurrent short-lasting human occupations testified by apparently undisturbed living floors with combustion features. These units yielded large quantities of macrofauna remains and lithics as well as microfauna and terrestrial and marine molluscs including several valves of Callista chione.

The lithic assemblage is still under study and we preliminary illustrate here only some technological characteristics of the artefacts. The raw materials are mainly flint, radiolarite and siliceous limestone deriving from small pebbles of local origin. At present, these pebbles can be collected in the Triassic formations cropping out close to the cave (Segre, 1959); however, the Neandertals may have collected this raw material also from now submerged alluvial deposits, like sea-shores and riverbeds located in the plains nearby the cave.

Lithic technology like knapping methods, production schemes and core convexity management still have to be investigated in depth. However, preliminary observations suggest that most of the lithic production of Grotta dei Santi can be associated with the Levallois concept. Some Levallois products suggest the use of the unipolar recurrent method aimed at obtaining elongated blanks. The volume of some cores was exploited according to the additional and non-integrated volumetric concept (Boeda, 2013) (Fig. 3). Interestingly these are cores aimed at the production of bladelets or small flakes.

Noteworthy is the occurrence of short production sequences, which are related to the small size of pebbles. Usually only a single generation of intended flakes followed the detachment of the first cortical flakes. After this, the core was already exhausted. This is the reason why most of the flakes have cortical parts on the dorsal face. The occurrence of use-wears on the edges resulting from the contact between the ventral and the cortical surfaces demonstrates that these parts, which usually represent the prehensive portions of the tools, could also be used as active edges owing to raw material constraints.

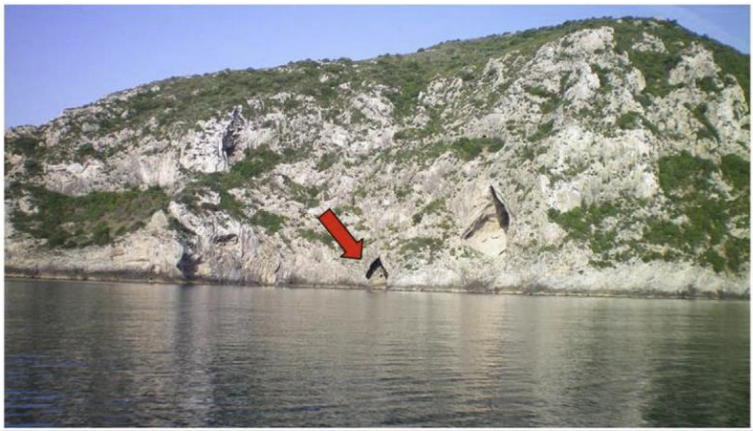
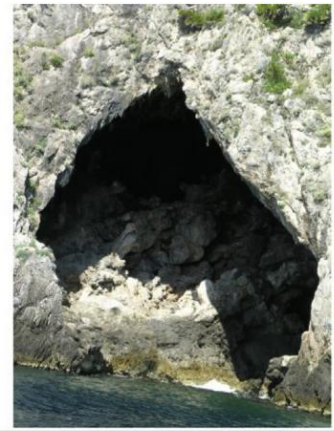
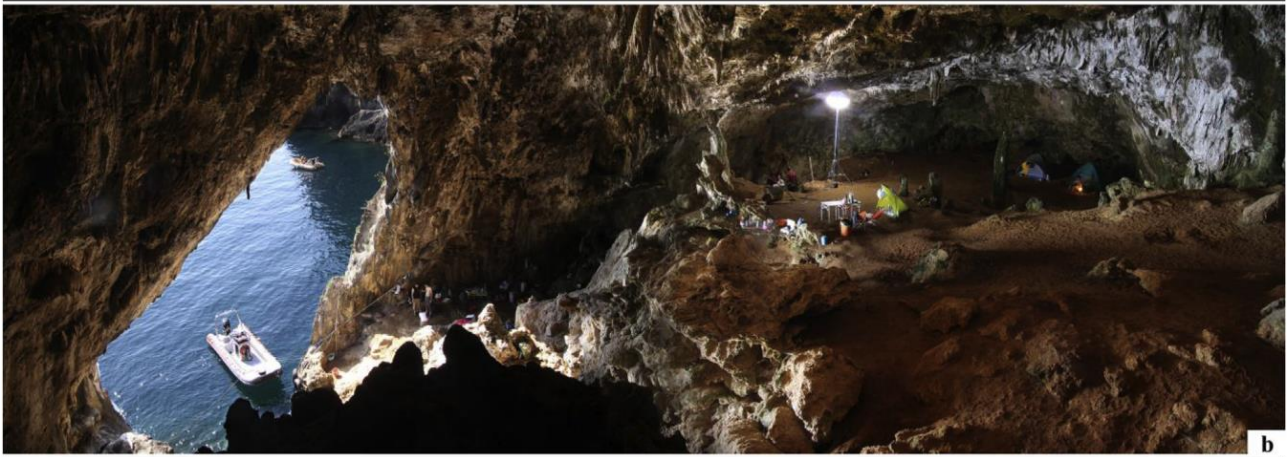
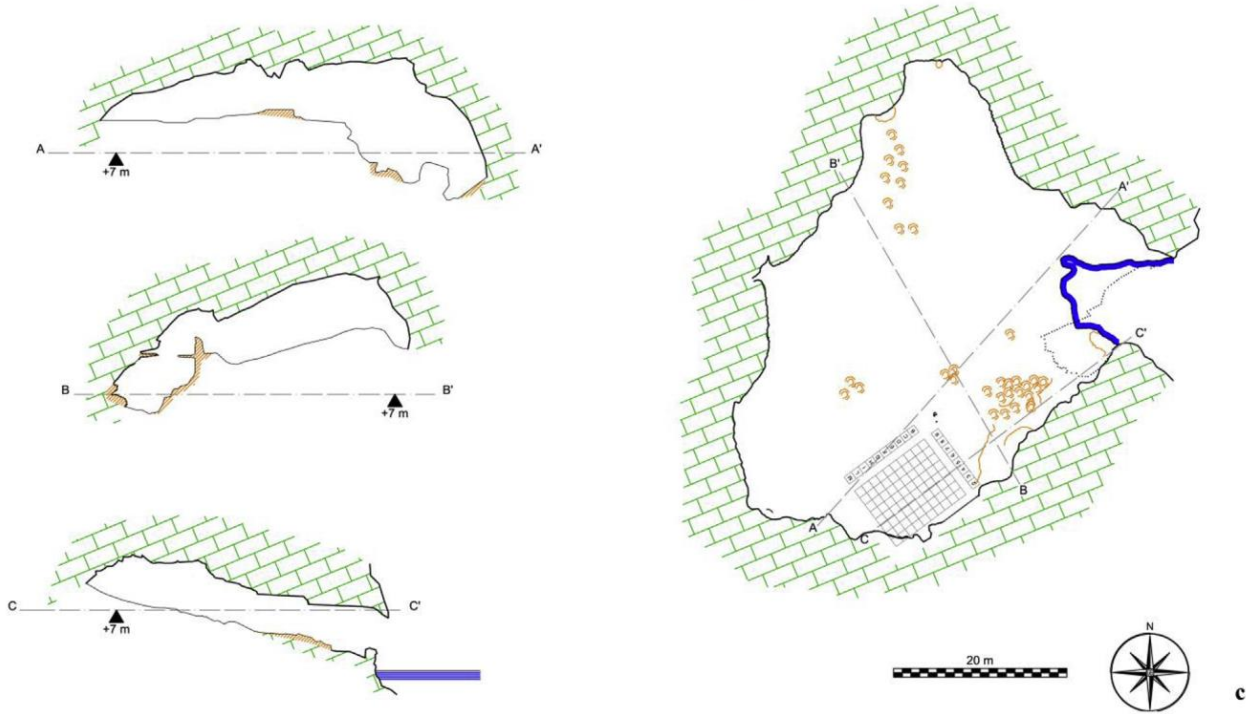

**Fig. 2.** General views and planimetry of the cave. a) Views of the cave entrance from East. The cave entrance is about 12m wide and opens inside a small cove resulting from the collapse of part of the cavity, which probably belonged to a much larger old underground system. Some niches e possibly parts of the same old system e that can be observed at different heights on the falaise, still include speleothem remains and finely layered sand-size clastic sediments. b) Panoramic view of the inside of the cave. c) Cross-sections and planimetry of the cave. Photos by Maurizio Lanfredini (a) and Stefano Ricci (b). Planimetry by Giovanni Boschian and Fabio Parenti.

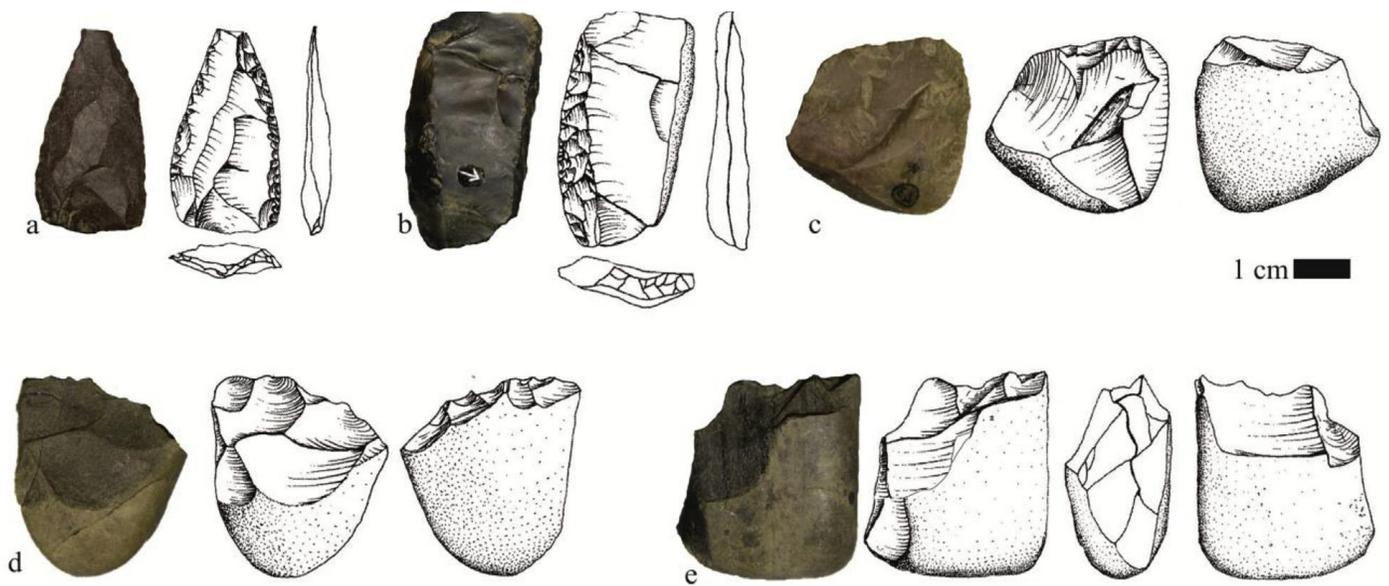

**Fig. 3.** Lithic industry of Grotta dei Santi. a) Retouched Levallois point; b) side scraper; c) d) e) additional cores. Photos by Vincenzo Spagnolo. Drawings by Giulia Marciani.

## 3. Materials and methods

### 3.1. Excavation techniques

Preliminary inspections and test excavations were carried out between 2007 and 2011 in order to plan systematic excavations. Extensive excavations were located in the southern "corridor" of the cave (Fig. 2), after all reworked sediments covering natural profiles and erosion surfaces were removed. A 1 1m reference grid was set by total station, and each square was further subdivided into four 50 50 cm sectors (I, II, III, IV). The anthropogenic units were excavated manually by trowels and brushes and, when necessary, by precision tools. Stratigraphic units thicker than 5 cm were further subdivided into technical spits thinner than 5 cm. All the excavated sediment was dry and wet sieved on a 0.5mm mesh and finally screened by tweezers on trays. Living floors and vertical profiles were mapped in 1:10 scale and photographed for 3D reconstruction. All remains visible at eye-scale were plotted in x, y, z coordinates.

### 3.2. Stratigraphic unit numbering

The upper part of the sequence was excavated following standard archaeological criteria from top downwards. Stratigraphic units were defined following a strictly lithologic criterion (North American Commission on Stratigraphic Nomenclature, 2005), by unit numbers starting from 100 and increasing downwards. Some stratigraphic units, or groups of units, were named by numbers much larger (500, 900, 1000 to 1004, etc.): these were located lower in the sequence than those under excavation at that time but had to be partially excavated in order to salvage them from erosion due to exceptional winter gales. In these cases, the naming was restarted from the aforementioned higher numbers in order to "leave space" for units still to be excavated.

Unlike the archaeological field procedure, the geoarchaeological study subdivided the sequence into stratigraphic units following the lithologic and allostratigraphic criteria (North American Commission on Stratigraphic Nomenclature, 2005). The naming of the units considers i) that the bottom of the sequence could be observed; ii) that units much older than the archaeological ones and iii) their subunits have to be included; iv) that the stratigraphic relationships between the outer (older) and inner (younger, including the archaeological units currently under excavation) parts of the sequence are still unclear. The units are consequently coded from the bottom upwards by groups of numbers separated by full stops (e.g. 10.1, 20.3.2e111): the first number differentiates the outer (10) from the inner (20) sequence and the following numbers indicate progressively lower rank units; the numbers after the dash indicate the archaeological units.

### 3.3. Geoarchaeology

The geoarchaeological study of the sequence was carried out partly on natural sediment outcrops, partly on profiles and surfaces excavated between 2008 and 2016. The deposit was divided into lithologic units observed at eye-scale and described in 3D during the excavations, beneficiating of extensive vertical profiles and areal exposure of the boundaries; the description is based on Catt (1991). The architecture of the units and their stratigraphic relationships were interpreted in relation with the geomorphologic characteristics of the cave. The geological characteristics of the surrounding area, which was repeatedly explored during the excavations, were used in reconstructing the sedimentary processes and the relationships between the cave infill and the outer landscape.

The geoarchaeological study reported in this paper is largely preliminary and tentative, because some units were explored only over small areas, some others were observed only over small extensions in natural outcrops, and because part of the stratigraphic relationships still needs elucidation. Geochronometric data are also still preliminary, and those presented in this paper concern only the most recent part of the sequence, whereas the older units are still undated. Consequently, the geoarchaeological work was focused on a description of the lithologic and architectural aspects of the sediments, on partial stratigraphic relationships, and on the interpretation of this evidence in terms of sedimentary models that may shed light on the site formation processes. Deeper geo-archaeological insights in climate change and human behavior will be provided after a wider set of chronological data is available.

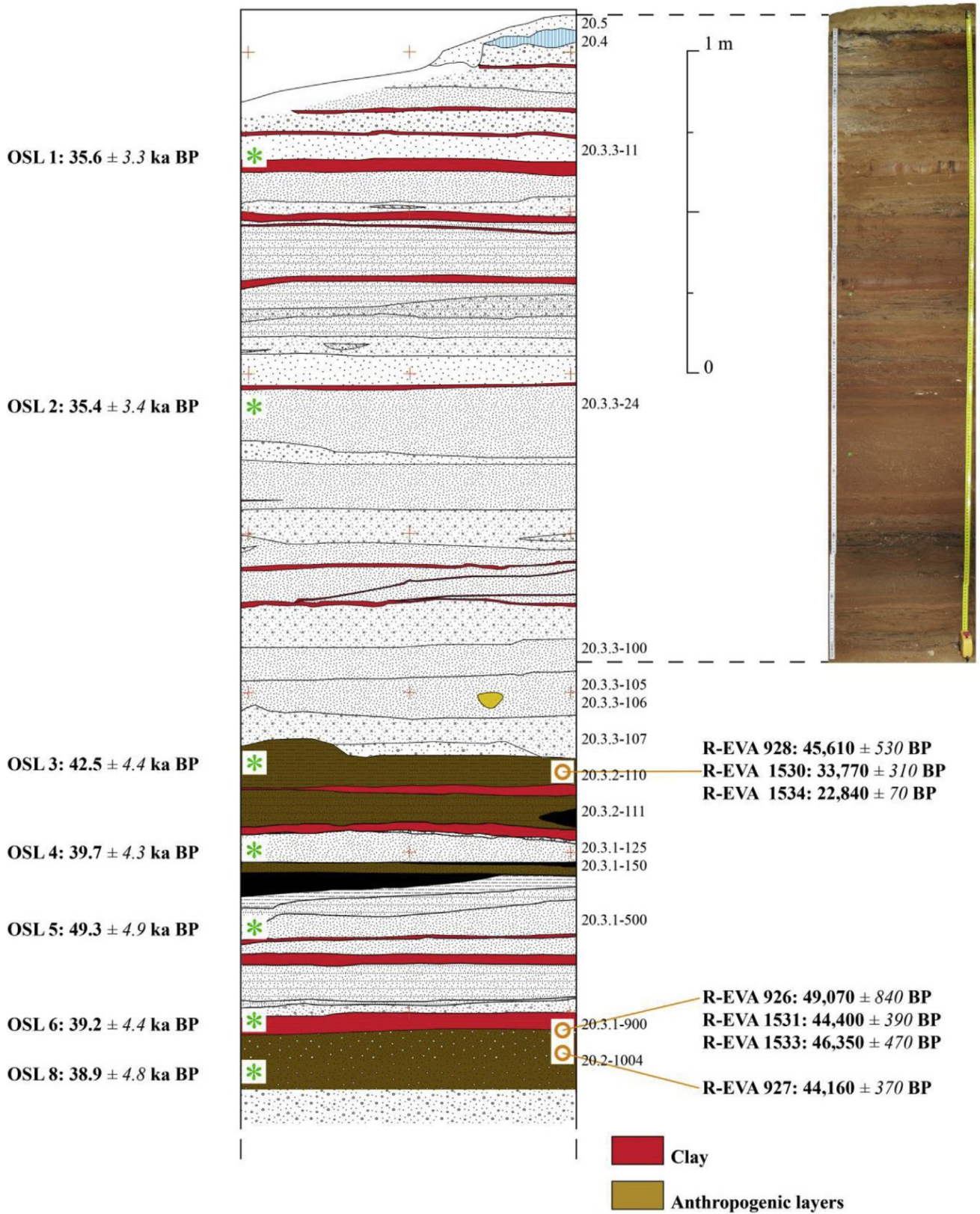

**Fig. 4.** Stratigraphic sketch of the cave investigated deposit (Fig. 5, panel C). OSL and 14C dates and the position of the samples taken for dating have been reported. The photo corresponds only to the upper portion of the drawing and the scale relates to both. Unit numbers are indicated to the right of the drawing. Black¼ hearths, yellow¼ coprolites, blu¼ stalagmite. Dotted texture varies according to the different granulometry of sands. Drawing by Vincenzo Spagnolo. (For interpretation of the references to colour in this figure legend, the reader is referred to the Web version of this article.)

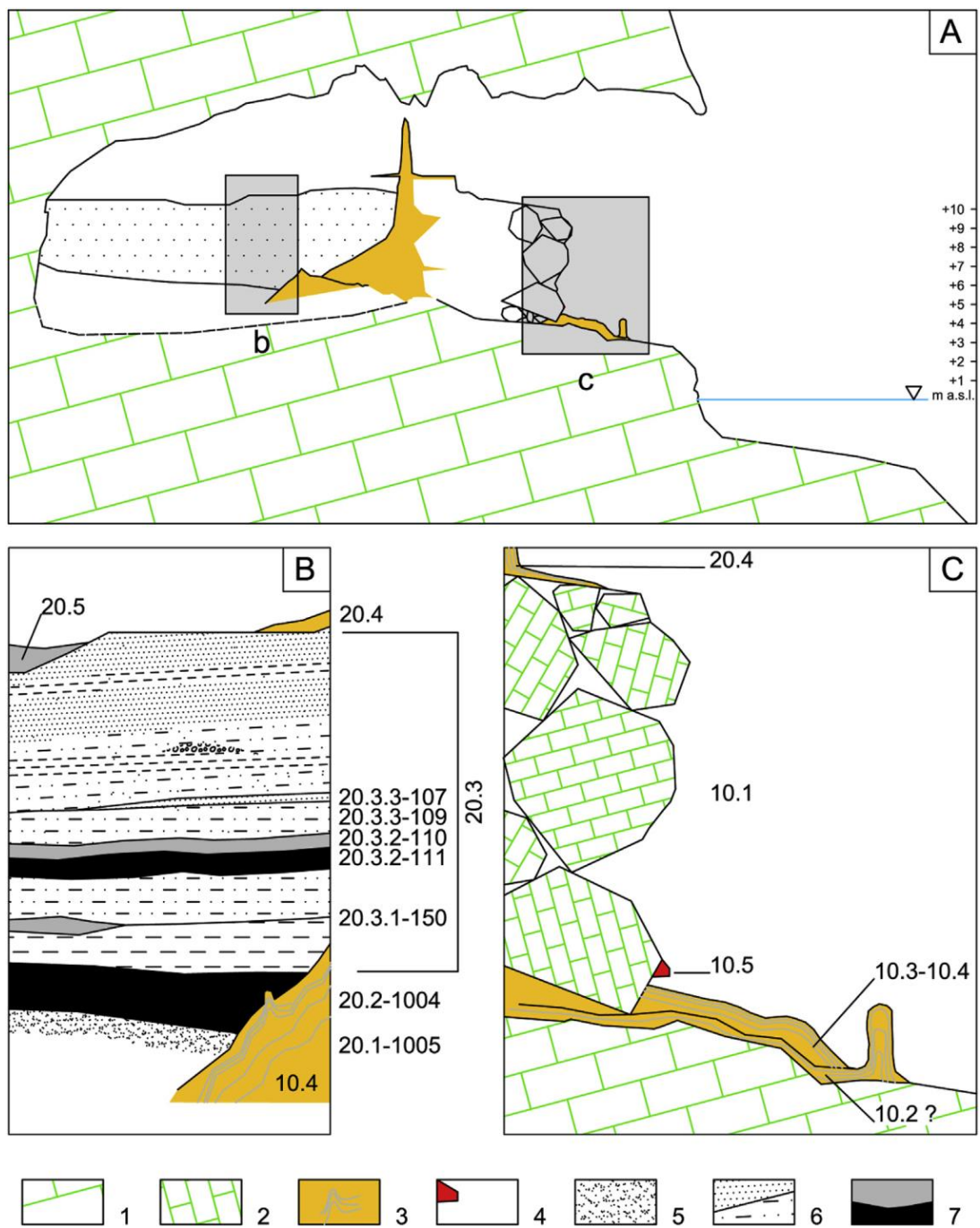

**Fig. 5.** Profiles of the Grotta dei Santi sequence. Panel A: longitudinal profile, showing heights above present-day sea-level and simplified architecture of the sedimentary units; b, c: sediment sequences represented in panels B (inner sequence) and C (outer sequence) respectively. 1: bedrock; 2: limestone blocks; 3: speleothems; 4: sandy marine sediments; 5: fine gravel with reworked mollusk shells; 6: interlayered sand to sandy silt loam to clay sediments, cyclically deposited; 7: anthropogenic sediments, gray shade intensity directly related to artifact/ash density. Panel B: inner sequence profile. Note that the stratigraphic relationships between units 10.1 and 10.2 are not clear: 10.2 is only tentatively represented as underlying 10.1, whereas 10.3 and 10.4 certainly overlie 10.1. Panel C: outer sequence profile. Unit 10.4 (flowstone) is at present the only stratigraphic unit occurring in both sequences. (For interpretation of the references to colour in this figure legend, the reader is referred to the Web version of this article.)

### 3.4. Macrofauna

We report the study of macromammal remains retrieved from units 20.3.2e110, 20.3.2e111, 20.3.1e150A and 20.2e1004 during the 2007e2014 excavations. The reference collection for taxonomy is the one stored at Dipartimento di Scienze Fisiche della Terra e dell'Ambiente, U.R. Preistoria e Antropologia, of the University of Siena. Age profiles were calculated by combining teeth eruption and wear, and epiphyseal fusion of the long bones (Grant, 1982). The unidentified specimens were divided into anatomical classes ("skull", "jaw", "teeth", "vertebrae", "ribs" etc., or more generally as "epiphysis", "diaphysis", and "fragments of cancellous bone") and into dimensional classes (1e3 cm, 3e6 cm, 6e10 cm, >10 cm); specimens freshly fractured during recovery were excluded from the count. The fragmentation degree of the bone assemblage was obtained from these data.

The taphonomic analysis was aimed at assessing the origin of surface modifications on the whole sample. Bone surfaces were studied using a HiroxKH-7700 3D digital microscope, that enables in-depth observations and both at low and high magnification (e.g. Arrighi et al., 2016; Oxilia et al., 2017); 3D images were obtained by the Auto Multi Focus tool, which stacks up to one hundred pictures shot at different focus levels. Metric parameters as defined by Bello and Soligo (2008) and Boschin and Crezzini (2012) were collected on amedian cross section of each identified cut-mark and carnivore tooth mark, as already tested in previous papers (Boschin and Crezzini, 2012; Crezzini et al., 2014; Moretti et al., 2015). Mann-Whitney U-tests and principal component analysis (PCA) were carried out in order to evaluate the metric parameters used in differentiating cut marks from tooth marks. Statistics were per-formed using the R software version 2.12.0 (© 2010 The R Foundation for Statistical Computing) and the Past software (Hammer et al., 2001).

### 3.5. Microfauna

The small mammal remains were retrieved from units 20.3.3e105, 20.3.3e106, 20.3.3e107, 20.3.2e110 and 20.3.2e111 and were collected, sieved, screened and stored during the 2007e2014 fieldwork. The identification was carried out using the reference collection of the University of Ferrara, Dipartimento di Studi Umanistici, Sezione di Scienze Preistoriche ed Antropologiche, following the general criteria indicated by Chaline (1972), Dupuis (1986), Felten et al. (1973), Niethammer and Krapp (1978) and Sevilla (1988). The remains were identified by a Leica M8 and a Wild M5 binocular microscopes, respectively at 4x and 25x magnification. The best diagnostic elements were used to identify the remains at species level: mandible, maxilla and isolated teeth for shrews; isolated teeth and humerus for Talpidae; isolated teeth for Erinaceidae; isolated teeth, humerus, mandible and maxilla for bats; first lowermolars for Arvicolinae;mandible and isolated teeth for Apodemus sp., Eliomys quercinus. The Grotta dei Santi sample includes a large number of remains belonging to the genus Apodemus that could not be identified at species level due to lack of teeth. However, this genus was identified by dental alveolus shape following Niethammer and Krapp (1978). All the remains belonging to the genus Apodemus (A. sylvaticus, A. cf. flavicollis and Apodemus sp.) were incorporated into the group Apodemus (Sylvaemus), because these species are considered sympatric and their combination would not affect the paleoenvironmental and paleoclimatic reconstruction.

The identified remains were successively grouped under minimum-number-of-individuals (MNI) by counting the most frequent diagnostic element in each macro-unit, taking into account the laterality.

The Habitat Weightings method was used in paleoenvironmental reconstruction (Andrews, 2006; Evans et al., 1981): considering that almost all Upper Pleistocene fossil small mammals are extant taxa, their environment can be determined by comparison with their present habitat. This method consists in distributing each small-mammal taxon in the habitat(s) where it can be currently found within the Italian Peninsula. Habitats are divided into six types: rocky (area with rocky or stony substratum), water (along streams, lakes and ponds), woodland (mature forest), open woodland (woodland margins and forest patches, with moderate ground cover), open dry (meadows under seasonal climate change) and open humid (evergreen meadow with dense pastures and suitable topsoil) (Lopez-García et al., 2014). The evenness of the small mammals community was calculated by the Simpson index of diversity (S¼ 1-Ppi2) where pi is the percentage of the individuals in the ith species on the total number of individuals (Berto et al., 2018; Harper and Hammer, 2006). The Simpson index is constrained between 0 and 1, with 0 corresponding to a community with a dominant taxon. Owing to the rather unexpected occurrence of a mostly southern taxon like Microtus (Terricola) savii at the site latitude, a biometric analysis was carried out following the criteria of Van der Meulen (1972) revised by Masini et al. (1997) in order to verify the attribution. Microtus (Terricola) savii teeth were photographed under a Leica EZ4 HD stereomicroscope and the program Image J was used to process the pictures and take measurement on first lower molars (Fig. 11). The Hernandez Fernandez (2005, 2001) methodwas applied to the paleoclimatic reconstruction; a value (1) was assigned to Microtus (Terricola) savii, - which is not included in the Hernandez Fernandez faunal list - in the climate category IV, which represents subtropical condition with rainy winters and dry summers (Berto et al., 2017).

### 3.6. Radiometric chronology

### 3.6.1. Radiocarbon dating

Bone and charcoal samples were collected from units 20.3.2e110, 20.3.2e111, 20.2e1004A and B (for the description of the units see paragraph 4.1). Unfortunately, no one of the 23 bone samples produced enough collagen for dating. The charcoal was sent directly to the Klaus-Tschira-AMS facility of the CurtEngelhorn Centre in Mannheim, Germany, where samples were pretreated using the ABOX method and the insoluble fraction was combusted. $CO_2$ was converted catalytically to graphite (Kromer et al., 2013).

### 3.6.2. Luminescence dating

Samples for luminescence dating were partly collected from sterile sandy or clay units intercalated within the anthropogenic units, and partly from units located above the anthropogenic series (Table 6; Fig. 4). Sample preparation of OSL samples included the common steps of chemical treatment (Fitzsimmons et al., 2014) including digestion in 10% HCl and in 15% $H_2O_2$ to remove carbonates and organic matter. The coarsegrain quartz fraction (180e250 mm) was isolated from heavy-minerals and feldspars using lithium heterotung state. Finally, the coarse quartz grains were etched using HF and then re-sieved to remove smaller grain-size fractions. All luminescence runs were conducted on a Risoe TL/DA-20 reader equipped with blue light-emitting diodes and acalibrated $^{90}Sr/^{90}Y$ beta source with a dose rate of about 0.24 Gy/s. To test for feldspar-contamination within the quartz-crystal lattice, sample material was additionally stimulated with IR-light emitting diodes (850± 50 nm). To detect the quartz luminescence signal a Hoya U-340 filter was used.

For luminescence measurements, 2mm sized steel discs were used. The usage of smaller aliquots or single-grains seemed not promising due to the relatively dim luminescence signal emitted by the quartz-grains under study. The preheat- and cutheat parameters for the SAR protocol (Murray andWintle, 2003) were set up by applying dose recovery tests to samples L-Eva 1452 and L-Eva 1454. Therefore, aliquots were bleached under a solar-lamp for 1 h. Subsequently, an artificial dose close to the expected natural one was inserted. In the following step, the SAR approach was used trying to recover the given dose. For DR tests, preheat-temperatures ranging from 150 to 250 C were tested and the cut heat was either set to 200 C or 220 C. For each temperature combination 4 aliquots were used.

Based on the results of the DR test, preheat- and cutheat temperatures were set to 180 C and 200 C respectively. The measured-to given dose ratios for that temperature-combination was at 0.97± 0.01 (L-Eva 1452) and 0.99± 0.03 (L-Eva 1454).

For final De-estimation, only aliquots showing and IR-depletion and recycling ratio deviating >10% from unity were included. The De-values quoted in Table 6 are based on the Central Age Model (Galbraith et al., 1999).

Regarding the Dosimetry, the concentrations of U, Th and 40K were measured on the dried sample material using low-level (high resolution) gamma-ray spectrometry. All gamma-measurements were conducted in the "Felsenkeller" laboratory in Dresden.

Additional to gamma-ray spectrometry in the laboratory, the gamma-dosimetry was measured in situ at the sampling points using $Al_2O_3$:C pellets (Richter et al., 2010), that were installed in the sedimentary sequence for about two years. Used dose rate conversion factors are based on Guerin et al. (2011).

## 4. Results

### 4.1. Stratigraphic sequence

A first description and interpretation of the cave sediments was provided by A.G. Segre (1959) after his brief visits in 1951 and 1954 (Table 1). The 2007e2016 excavations explored only the most recent part of the sequence (A.G. Segre units a1, a2 and d) which crops out in the southern channel and includes the bulk of the cultural remains. The older units e st1 to br2, some including faunal and allegedly also cultural remains e were only observed on natural outcrops in the outer part of the cave. Not surprisingly, the results of our observations do not match exactly those of A.G. Segre, as the outcrops may have changed after some sixty years of exposure to the waves of the southerly winter gales. To avoid being entangled in previous perspectives and interpretations, our study re-examines from scratch the sequence and results in a novel appraisal of the stratigraphic sequence, including some units that could not be observed by A.G. Segre during his brief visits. The supposed match with A.G Segre observations is reported in Table 1.

The cave floor is shaped in the "Calcare Cavernoso" limestone formation; it is exposed only in the outermost part of the cave, where wave erosion shaped it into a step about 2.5m high above the present-day sea-level. This is the lowest observed height of the cave bedrock, which was probably shaped into a marine erosion platform gently climbing towards the inside of the cave. Unfortunately, the real height and extension of the platform cannot be estimated because of strong marine erosion and karstification, and because all its inner part is covered by sediments.

The sequence of sediments that infill the cave can be divided into two main parts (Fig. 5) differing by age and lithological characteristics. The outer one (Fig. 5, C) can be observed along a marine erosion scarp that faces the sea and overlies the step of the cave bedrock; its bottom rests upon the cave bedrock. The inner sequence (Fig. 5, B) crops out well inside the cave, along the side of the "corridor" that rims the southern wall of the cave. The bottom of this sequence could not be observed along all the outcrop, so that its whole thickness is still unknown; hypotheses about its vertical and lateral extent mostly derive from surface observations, as well as from some sediment augering carried out within the cave.

| (Segre, 1959) | | | (this paper) |
|---|---|---|---|
| Unit | Unit description | Chronostratigraphy | Unit number |
| e | Yellow sands including Copper Age hearths | Holocene | 20.5 |
| $st_3$ | Upper stalagmite | Würm 3 | 20.4 |
| $a_2, a_1$ | Red clay horizons including *Caecilianella acicula* remains; overlain by a thin detrital layer | | not observed |
| d | Climbing dune; reddish sands, yellowish at the top, and with cemented crusts at the base | | 20.3.1–150 to 20.3.3–105 |
| $br_2$ | Upper bone breccia, interlayered with cemented red silt | Würm 2 | 20.2–1004 |
| $st_2$ | Thick lower stalagmite | Würm 1 | 10.4–10.3 |
| $br_1$ | Breccia with Ibex remains and Mousterian industry | Anawürm 1 | not observed |
| t | Shore deposits | Tyrrhenian | 10.5 |
| fr | Blocks - Ceiling breakdown | | 10.1 |
| $st_1$ | Stalagmite remains | Pre-Tyrrhenian | 10.2 (?) |

**Table 1:** Field summary description and chronostratigraphic assessment of the Grotta dei Santi sequence, as reported by A.G. Segre (1959: Fig. 3, p. 6). Units are listed following stratigraphic order. Correspondence with the units defined by this paper is reported in the last column.

The most outstanding unit of the outer sequence is a 4-6m- thick scree-like accumulation of very large (up to 3e4m wide) limestone blocks (unit 10.1) that occupies approximately the middle of the cave and tapers to the inside. The outer part of this unit was cut by marine erosion into the subvertical scarp facing the sea.

It is at present unclear whether these blocks lie directly upon the cave bottom and represent the base unit of the clastic infill, or if other units can be found beneath the blocks. In fact, a discontinuous flowstone (unit 10.2) and some remains of stalagmite bosses grow directly upon the surface of the basal limestone in the area between the scarp and the limestone step, but it is not clear if the flowstone underlies the blocks or just overlies them.

Following A.G. Segre (1959), a stalagmite level (st1) should underlie the blocks and consequently correspond to 10.2, but its stratigraphic position is unclear and appears rather hypothetically from his field drawings; it may alternatively correspond to a thin flowstone (unit 10.3) that covers the floor of the southern corridor. Unfortunately, the stratigraphic situation is complicated by several other younger discontinuous flowstones (unit 10.4) that mantle the blocks and dip steeply into the corridor, where they become horizontal and roughly parallel to flowstone 10.3. These other crusts cover discontinuously the blocks and locally grow intowide and tall stalagmites.

Flowstone 10.4 dips rather steeply southwards and westwards from the top of unit 10.1 into the inside of the cave. All the other units of the inner sequence observed till today lie upon 10.4, which is at present the only unit occurring in both sequences and is consequently the only stratigraphic connection between the two sequences.

Remains of cemented sandy marine sediments (unit 10.5) including some fine gravel, shell, coral and bryozoan fragments adhere to the blocks and to the rock walls (but not to the speleothems), up to about 4e4.5m a.s.l. These sediments are characteristic of backshore/foreshore environments and were clearly deposited after the accumulation of the blocks and the start of the scarp erosion.

The lower part of the inner sequence is still unexcavated, whereas the upper one lies unconformally upon a flowstone, which can be tentatively correlated with unit 10.4. This flowstone dips westwards and southwards forming a depression filled by the inner sequence. To the outside, the depression is barred at about 7.2m a.s.l. by a ridge of stalagmite bosses running perpendicular to the southern wall.

At present, the lowermost observed unit of the inner sequence is a reddish loamy sand with cm-size rounded pebbles and reworked mollusc shell fragments, including also larval forms (unit 20.1), which crops out in a small area (about 2m2) recently exposed by marine erosion. The limit with the overlying unit 20.2e1004 is marked by the top of some limestone blocks lined approximately at the same level.

Unit 20.2-1004 is in fact a complex sequence of at least two e but probably more - blackish horizons rich in organic matter (units 20.2.4-1004A and 20.2.2-1004B), alternating with layers of greyish ash and of rubified sediment. The thickness of these subunits varies between 3-4 and 10-12 cm.

Unit 20.2-1004 is overlain by 3m of finely alternating greyish sand and reddish clay loam (unit 20.3), starting with a 10-15 cm- thick horizon of red clay. These sediments are organised in fining-upwards cyclical sequences organised as follows:

- Medium (sometimes fine) to coarse greyish sand, with frequent small subangular to subrounded pebbles and granules. Lenticular bedding and wave ripples, sometimes with flaser bedding are the dominant sedimentary structures; femic volcanic minerals (mostly euhedral to anhedral pyroxene, [Fig. 6]), and/or small pebbles can occur within the concavities. The top of this subunits can sometimes be more evenly laminated. The bulk of the sand grains are K-feldspar and quartz, with subordinate other minerals. Thickness 8e12 cm, sharp erosional limit.

- Fine to medium reddish silty sand, usually massive or with few laminae (up to 1 cm thick) of reddish to yellowish clay. Thickness 6-10 cm, abrupt to clear limit.

- Massive red clay, without sedimentary structures. Thickness 2-10 cm, abrupt limit.

The rounded sand grains surface pattern is frozen, whereas the subangular ones' is matte, indicating aeolian transport of the mature grains and river transport of the fresh ones.

Layers with more pebbles are mostly frequent in the upper part of the unit, and in the central part of the cave.

All the aforementioned levels are prevalently parallel and sub-horizontal. In the bottom area of the cave, they dip very gently towards the bottomwall, where the sequences are more articulated and include more sublevels. Several minor erosion surfaces complicate the sequence, mostly within the lowermost part of the unit.

Human and/or animal inputs occur in the following subunits of unit 20.3.

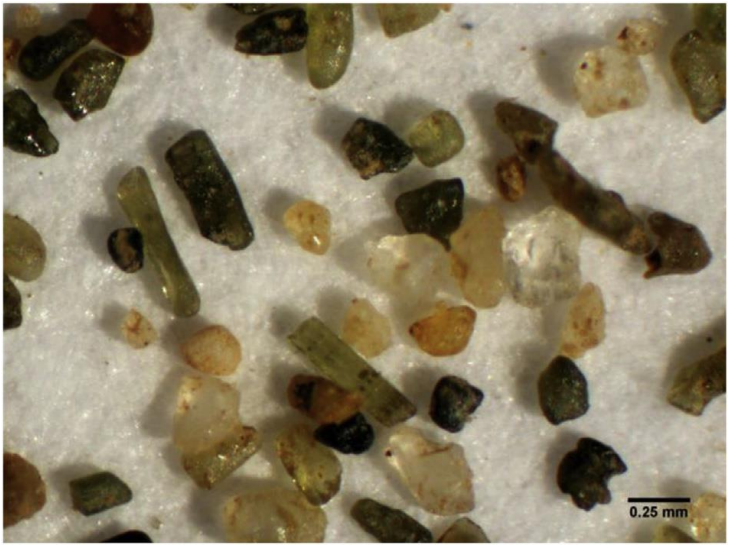

**Fig. 6.** Unit 20.3, sandy levels of Grotta dei Santi. Washed 125e250 mm grain-size fraction. Note autoptically identifiable subhedral/euhedral (sometimes slightly rounded) elongated hexagonal prisms of green pyroxene; anhedral, equant angular uncoloured hyaline transparent K-felspar (sanidine); anhedral, equant yellowish to milky white subrounded quartz; anhedral, equant subrounded to subangular dark gray to opaque black magnetite. Grain surfaces are often frozen or matte, indicating river and aeolian transport. (For interpretation of the references to colour in this figure legend, the reader is referred to the Web version of this article.)

Unit 20.3.1-150 is a 10-15 cm-thick layer of laminated sand containing at least three anthropogenic levels with hearths and Mousterian industry.

Unit 20.3.2-111-110 are two respectively 15 cm and 20 cm-thick levels of laminated sand. They include sublevels with lithics and fauna, alternating to non-cultural levels with sparse phosphate lenses and coprolites, whose shape and size are compatible with spotted hyena (Crocuta crocuta spelaea) (Crezzini and Moroni, 2012). No charcoal was found within two hearths situated to the inside of the cave. Mousterian stone tools are common in the anthropogenic levels.

Unit 20.3.3-109-107-106-105 shares the lithological characteristics of the underlying sublevels and includes frequent hyena and lion coprolites. Lenses of whitish amorphous phosphates (up to 2e3 cm thick, several decimetres wide) located at different heights within the unit derive from coprolite trampling and weathering. No trace of human activity is documented.

The top of unit 20.3 is locally covered by stalagmites and discontinuous flowstone (unit 20.4) growing in the same spots as the previous ones. These flowstones form some false floors along the cave walls, most evidently in the southern "corridor".

Unit 20.5 is represented by discontinuous patches of mollusk shell-rich sediment and small hearths of Roman age, which partly reworked the underlying sediments.

Several breccia remains - often including bone fragments - adhere to the cave bottom and walls in various positions and at different height. At present, any correlation with the main sequence is problematic, but they indicate that the cave infill was more extended towards the outside in the past. A patch of terra rossa with fauna and lithics (possibly breccia br2 [Segre, 1959]) is located near the southern wall, about 20m to the inside of the cave. Unfortunately, this outcrop is not more than 3m2 wide and cannot be correlated with the main sequence, even if it lies upon a flowstone that may correspond to unit 10.4.

### 4.2. Cultural sequence

Thorough profile cleansing and extensive excavations carried out during the 2011e2013 field seasons uncovered a set of previously unobserved horizons rich in faunal remains (unit 20.3.2e110) and Mousterian lithics (unit 20.3.2e111 and 20.3.1e150), all included in the unit (20.3) overlying the lowermost Mousterian units (20.2.4e1004). The following cultural macro-units were identified, from bottom upwards (Fig. 5):

D) The lowermost macro-unit (20.2e1004) was excavated only in an area slightly larger than 1m2 that was threatened by wave erosion. This level includes a still unknown number of interfingered horizons including living floors and juxtaposed hearths rich in charcoal. Apparently, these hearths vary considerably in size e probably from 20 cm up to more than 1m e and in thickness. Unfortunately, units from 20.3.1 to 150 to 20.3.1e111 are extensively damaged by a several meters long tunnel, possibly a badger den.

C) Units 20.3.1e150 includes at least three closely spaced living floors. The upper one (20.3.1e150A) yielded lithics, several faunal remains and malacofauna, along with a hearth that had been partially destroyed by sea erosion. Rather interestingly, valves of the marine mollusc Callista chione, occasionally used as blanks for Mousterian tool production in other Italian coastal cave-sites (Borzatti Von Lowenstern, 1965; Palma di Cesnola, 1965; Romagnoli et al., 2015, 2016; Vicino, 1972) occur in this unit. However, all the specimens recovered so far at Grotta dei Santi are unmodified.

B) Mousterian cultural remains were collected over an area of 9m2 in units 20.3.2.111 and 20.3.2.110, within anthropogenic horizons interbedded with non-cultural levels including variously preserved coprolites. These levels are overlain by 2m of sterile sediments (unit 20.3), including coprolite-rich horizons at the bottom (units 20.3.3-107, 20.3.3-106 and 20.3.3-105).

Several stone artefacts, mammalofauna and malacofauna remains were found in unit 20.3.2e111, in association with two hearths. One of these (Hearth H6111 spit 2) e largely destroyed by a burrower tunnel e was about 60 cm wide and 5e7 cm thick. The second one (Hearth I5 111 spit 2) e still partially unexcavated e is well preserved and several artefacts and bones were found clustered around its edge. It is about 70 cm wide and 10 cm thick. These small combustion features are simple thin ash layers overlying burnt reddish sediment, without any charcoal, significantly differing from the hearths occurring in macro-unit D, which comprise of alternating ash and charcoal-rich layers.

A) The Holocenematerial was found in some test trenches in the top 15e20 cm of the sequence. It is represented by sparse Roman Age hearths and pottery, discontinuously distributed throughout the inner part of the cave.

### 4.3. Macrofauna

The macromammal assemblages from units 20.2-1004, 20.3.1-150A, 20.3.2-111 and 20.3.2-110 include 228 identified specimens (Table 2) characterised by a high frequency of cervids. The most abundant ungulate was Cervus elaphus, followed by Dama dama (absent in 20.3.2e111), Capreolus capreolus (not occurring in 20.3.1-150A) and Bos primigenius.

The combination of all age-at-death data clearly indicates that the majority of red deer and other cervids in units 20.3.2e110 and 20.3.2e111 are adult individuals, whereas the juvenile and old ones are less represented. Stephanorhinus hemitoechus (Pandolfi et al., 2017) occurs in unit 20.3.1-150A, and two large carnivores, Crocuta crocuta spelaea and Panthera pardus, also represented by a large amount of coprolites, occur in 20.3.2-110. Capra ibex occurs only in unit 20.2e1004. Equus ferus occurs in units 20.3.1-150A and 20.3.2-110. A climate change probably occurred between units 20.3.1e150A and 20.3.2-110, as indicated by a decrease in forest related taxa (i.e. cervids) and an increase in forest steppe – or steppe-related taxa (Bos primigenius, Equus ferus, Stephanorhinus hemitoechus).

| Units | 20.2–1004 | | | | 20.3.1–150A | | | | | | | 20.3.2–111 | | | 20.3.2–110 | | | | | | | |
|---|---|---|---|---|---|---|---|---|---|---|---|---|---|---|---|---|---|---|---|---|---|---|
|  | Bp | Ci | Ce | Vv | Bp | B/B | C | Ce | Dd | Ef | S | Bp | Cc | Ce | Bp | Cc | C | Ce | Cr | Dd | Ef | Pp | su |
| Ant./Horn |  | 1 | 1 | 1 | 3 | 1 | 1 | 3 | 1 | 1 | 1 | 1 | 1 | 2 | 1 | 2 |  | 2 | 1 | 1 | 1 | 1 |  |
| Skull |  | 1 |  |  |  |  |  | 1 |  |  |  |  |  | 2 |  |  |  |  |  |  |  |  |  |
| Max |  |  | 1 |  |  |  |  | 1 |  |  |  |  |  |  |  |  |  | 2 |  |  |  |  |  |
| Teeth | 1 |  | 1 |  | 5 | 2 |  | 13 | 5 | 1 | 1 |  | 2 | 3 | 1 | 5 |  | 18 | 1 | 1 |  | 2 |  |
| Mand |  |  |  |  | 2 |  |  | 2 |  | 1 |  |  | 1 |  |  | 2 |  | 3 |  | 1 |  |  |  |
| Vert |  |  | 1 |  |  |  |  |  |  |  |  |  |  |  |  |  |  |  |  |  |  | 1 |  |
| Ribs |  |  |  |  |  |  |  |  |  |  |  |  |  |  |  |  |  |  |  |  |  |  |  |
| Ster |  |  |  |  |  |  |  |  |  |  |  |  |  |  |  |  |  |  |  |  |  |  |  |
| Sc |  |  |  |  |  |  |  | 1 |  |  |  | 1 |  |  |  |  |  |  |  |  |  |  |  |
| Hum |  |  |  |  | 2 |  | 1 | 2 |  |  |  |  |  |  | 2 |  |  | 1 |  |  |  |  |  |
| Rad |  |  |  |  | 1 |  |  | 1 |  | 1 |  |  |  | 1 |  |  |  |  | 1 |  | 1 |  |  |
| Ulna |  |  |  |  | 1 |  |  |  |  |  |  |  |  | 1 |  |  |  | 1 | 1 |  |  | 1 |  |
| Rad.-Ul |  |  |  |  | 1 |  |  | 1 |  |  |  |  |  |  |  |  |  |  |  |  |  |  |  |
| Mc |  |  |  |  |  |  | 1 | 1 |  |  |  |  |  | 2 |  |  |  | 3 |  |  |  | 2 |  |
| Carp |  |  |  |  |  |  |  |  |  |  |  |  |  | 1 |  |  |  | 2 |  |  |  | 1 |  |
| Inn |  |  |  |  |  |  |  | 2 |  |  |  |  |  |  |  |  |  |  |  |  |  |  |  |
| Fem |  |  |  |  |  |  |  |  |  |  |  |  |  | 1 |  |  |  | 4 |  |  |  |  |  |
| Tib |  |  |  |  | 6 |  |  | 2 |  |  |  |  |  |  | 1 |  |  | 5 |  |  |  |  | 1 |
| Mt |  |  |  |  | 1 |  | 1 | 6 |  |  |  |  |  | 3 |  |  |  | 8 |  |  |  | 2 |  |
| Tar |  |  |  |  |  |  |  | 1 |  |  |  |  |  | 1 |  |  |  | 2 |  |  |  |  |  |
| Ses |  |  |  |  |  |  |  |  |  |  |  |  |  | 2 |  |  |  | 2 |  |  |  |  |  |
| Mp |  |  |  |  |  | 1 |  |  |  |  |  |  |  |  | 1 |  | 1 | 1 |  |  |  |  |  |
| Ph |  |  |  |  |  |  | 1 | 3 |  | 1 |  | 1 | 9 |  | 1 | 2 |  | 5 |  |  |  | 3 |  |
| TOT NISP | 1 | 2 | 3 | 2 | 22 | 4 | 5 | 40 | 6 | 5 | 2 | 2 | 5 | 28 | 7 | 11 | 1 | 59 | 4 | 3 | 2 | 13 | 1 |
| % | 20 | 20 | 40 | 20 | 26.2 | 4.8 | 5.9 | 47.6 | 7.1 | 5.9 | 2.5 | 3.2 | 13 | 83.8 | 6.5 | 9.8 | 1.1 | 61.9 | 3.3 | 2.2 | 1.1 | 13 | 1.1 |

**Table 2:** Faunal associations according to units and skeletal districts. Abbreviations: Bp: Bos primig elaphus; Dd: Dama dama; Ef: Equus ferus; S: Stephanorhinus sp.; su: small-size ungulate; Cr Max: Maxilla; Mand: Mandible; Vert: Vertebra; Ster: Sternum; Sc: Scapula; Hum: Humeru Femur; Tib: Tibia; Mt: Metatarsal; Tar: Tarsal bones; Ses: Sesamoidal bones; Mp: Metap

Taphonomic analyses carried out on ungulate bones of units 20.3.2e110 and 20.3.2e111 reveal frequent anthropogenic traces: striae, percussion marks and cones. The cross-sections of striae recorded on identified samples by 3D microscopy are shallow, symmetric and V-shaped. While archaeological evidence (coprolites, large carnivore skeletal elements) (Moroni et al., 2010) points to the presence of carnivores in the cave, most of the striae do not show the shallow and broad U-section characterising tooth marks. The shape of the cross-sections does not look compatible with carnivore-modification and the origin of these grooves can be consequently related to butchery.

Additional 3D microscopy data were obtained by comparing archaeological evidence with morphometric data of experimental cut marks (Boschin and Crezzini, 2012) and tooth marks (Duches et al., 2016). A principal component analysis (PCA) was performed in order to evaluate these data and differentiate experimental cut marks from tooth marks. Within the 2D results (Fig. 7a) representing approximately 89% of the sample variability, the striae of Grotta dei Santi overlap with the modern cut marks.

PC1 accounts for 64.5% of the sample's variability and is a function of the size of the grooves. Considering this parameter, the striae of Grotta dei Santi are different both from tooth marks (Mann-Whitney U test p¼ 6.14E-08) and experimental cut marks (Mann-Whitney U test p¼ 5.887E-05). PC2 is a function of the shape of the striae and differentiates between the more U-shaped enius; tooth marks and the striae from Grotta dei Santi (p¼ 0.0002) but not between the latter and the experimental cut marks (p¼ 0.07). Fig. 7b shows the ratio between breadth at the top and breadth at the floor of the striae (RTF, a function of the groove shape, Boschin and Crezzini, 2012) for the two sets of cut-marks and the experimental tooth marks, confirming results of the PCA.

Small fragments (1e3 and 3e6 cm, 75% and 17% respectively) dominate among the unidentifiable bone remains recovered from unit 20.3.2e110, probably because of the widespread practice of bone fragmentation carried out by Palaeolithic people. Skulls (maxillary fragments, mostly isolated teeth and jaws) and limbs of slaughtered ungulates were the parts most frequently transported to the sites, regardless of the animal size, whereas unidentified fragments of the axial skeleton (vertebrae and ribs) are evidently scarce.

Cutting tool marks are the most frequent anthropogenic traces within the unidentified sample. These traces are mainly located on shaft and rib fragments and are probably related to carcass defleshing. Similarly to cut marks, impact traces and green-bone fractures are mainly located on limb elements of the identifiable set, and on diaphysis remains of the unidentifiable one. In both cases, these traces are related to long bone fracturing for marrow extraction. The fracture profiles are curved (Villa and Mahieu, 1991, Fig. 7b) with smooth edges (Morlan, 1984; Villa and Mahieu, 1991), and the angle formed by the cortical bone surface is oblique (Johnson, 1985; Morlan, 1984; Villa and Mahieu, 1991), suggesting that the bones were fractured when still fresh.

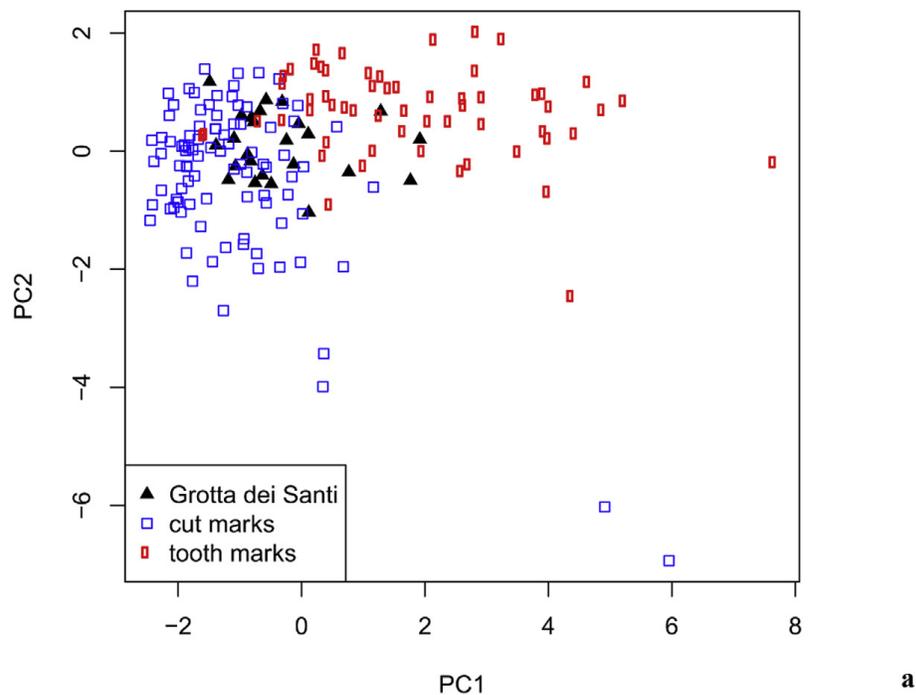

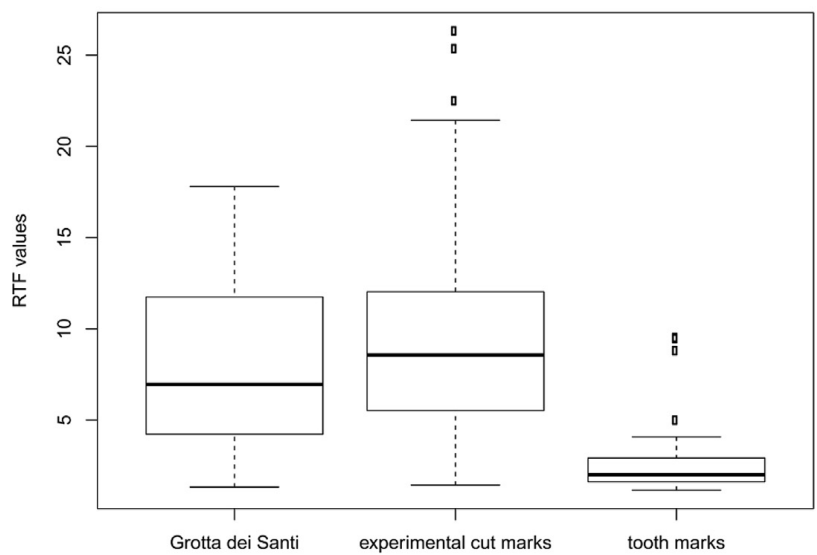

**Fig. 7.** Grotta dei Santi macrofaunal analyses. Principal component analysis of the measurements taken on the striae (a) and differences in the ratio between the breadths at the top
and the breadths at the floor (b). a) Within the 2D results representing approximately 89% of the sample variability, the striae of Grotta dei Santi overlap with the modern cut marks.
b) The figure shows the ratios between breadth at the top and breadth at the floor of the striae (RTF, a function of the groove shape, Boschin and Crezzini, 2012) for the two sets of
cut-marks and the experimental tooth marks.

### 4.4. Microfauna

The fossil assemblage comprises of 891 identified specimens corresponding to a minimum number of 448

individuals, belonging to at least 14 species. In order to obtain more reliable results, only the unit with enough individuals (20.3.3e107) was considered singularly, whereas the other ones were combined into twomacro units (unit 20.3.3e105 with 20.3.3e105; unit 20.3.2e110 with 20.3.2e111) including a total of 653 remains corresponding to a minimum number of 254 individuals.

The small mammals assemblage includes seven rodent taxa, three insectivores and five bats (Table 3), including the northern-most Late Pleistocene occurrence of Microtus (Terricola) savii. Within the whole assemblage, the dominant taxa are Apodemus (Sylvaemus), Microtus arvalis and Arvicola amphibius. A high percentage of bats was observed, particularly in macro-unit 20.3.2-110-111, which yielded 127 remains of Nyctalus noctula (Fig. 8).

Macro-unit 20.3.2-110-111 is the richest of the sequence with 455 remains corresponding to 160 MNI; it is dominated by Apodemus (Sylvaemus), together with Eliomys quercinus and Nyctalus noctula, which are species correlated to forest habitats. Unit 20.3.3e107 shows almost the same percentages as 20.3.2-110-111, except for a strong reduction of Nyctalus noctula. A change in habitat percentages can be observed between the units group 20.3.2-110-111 and 20.3.3e107, and the overlying macro-unit 20.3.3-105-106 (Fig. 9). Macro-unit 20.3.3-105-106 is dominated byMicrotus arvalis (45% of the assemblage) and also includes 12% of remains belonging to the group Apodemus (Sylvaemus), testifying to open-dry (OD) environments.

Following the Habitat Weighting Method, this transition corresponds to a reduction of Woodland (Wo) linked to the decline of Apodemus (Sylvaemus), and to an expansion of the Open Dry (OD) habitat related to the high percentage of Microtus arvalis. The habitat categoryWater maintains almost the same percentage in all three macro-units. From unit 20.3.3e107 to 20.3.3-105-106 biodiversity decreases, as showed by the Simpson index (Fig.10). Climate reconstruction following the Hernandez Fernandez method (Fig. 10) highlights a fluctuation in the sequence, which also coincides with the transition from unit 20.3.3e107 to 20.3.3-105-106, ndicating a decrease in average annual temperature, warmest month maximum and coldest month minimum temperature and annual precipitation.

Finally, measurement carried out on supposed Microtus (Terricola) savii specimens (Table 4) confirmed the attribution to this taxon, allowing a comparison between the assemblage of Grotta dei Santi and those of other MIS3 Italian sites (Petruso et al., 2011) (Fig. 11).

| | 20.3.2-110-111 | 20.3.3−107 | 20.3.3-105-106 |
|---|---|---|---|
| *Eliomys quercinus* | 8 4,91% | 3 4,48% | |
| *Apodemus (Sylvaemus)* | 42,77% | 13 22,39% | 4 12,12% |
| *Arvicola amphibius* | 16 9,82% | 15 28,36% | 4 12,12% |
| *Microtus agrestis* | 2 1,23% | 1 1,49% | 2 6,06% |
| *Microtus arvalis* | 26 15,95% | 12 17,91% | 15 45,45% |
| *Micortus (Terricola) savii* | 4 4,29% | 1 1,49% | 2 9,09% |
| *Erinaceus europaeus* | 1 0,61% | | |
| *Sorex* ex gr. *Araneus* | 2 1,23% | | |
| *Talpa europaea* | 1 0,61% | 1 1,49% | 3 6,06% |
| *Rhinolophus* gr. *euryale-mehelyi* | 8 4,91% | 3 4,48% | 1 3,03% |
| *Rhinolophus ferrumequinum* | 3 1,84% | 2 2,99% | |
| *Nyctalus noctula* | 32 19,63% | 5 7,25% | 2 6,06% |
| *Myotis* sp. | 12 7,36% | 2 2,99% | |
| *Miniopterus schreibersii* | 3 1,84% | 3 4,48% | |
| **Total NMI** | **160** | **61** | **33** |

Table 3: Grotta dei Santi, Small mammals assemblage, MNI, and percentage of the three macro-unit.

## 4.5. Radiometric chronology

The radiocarbon results are given in Table 5. The samples R-EVA 926 and 928 produced older ages hence are not consistent with the stratigraphy and archaeological evidence, this is due to the fact that they have been recovered from flotation with sea water. For this reason, we consider them 100% outliers in the Bayesian model. Other two samples R-EVA1534 and 1530 do not fit with the chronological situation produced with the other 14C dates as well as with the OSL. These samples are coming from an area that was subject to intense dripping from the ceiling, hence needs to be studied in more detail in the future. For this reason, we will exclude them from the Bayesian model.

The OSL ages of the sterile sediments covering the anthropogenic units yield a minimum age for the archaeological findings. They are in total agreement to each other and might point to a relatively quick deposition of that sediment-unit. A more precise chronological resolution is not possible keeping in mind the errorrange of the luminescence ages.

The ages from units containing artefacts point to a deposition of the sediments around 40 ka. Only sample OSL 5 yields a luminescence age of around 50 ka and is in disagreement with the quartz OSL ages from the over- and underlying sediment-units.

The Bayesian model, incorporating both 14C and OSL dates (Fig. 12), was produced using the OxCal 4.3 program (Bronk Ramsey, 2009) and the latest International Calibration Curve, IntCal13 (Reimer et al., 2013) and each OSL age was inserted as a C Date in calendar years before 1950 using 2s error. This allows to give a rough estimation of the starting and ending of the different units represented by the different boundaries (Table 7). More chronological work will be done in the future on Grotta dei Santi.

In the model was computed a General t-type Outlier Model (Bronk Ramsey, 2009) to detect problematic samples with prior probabilities set at 5% except for the 2 samples discussed above (R-EVA 926 and 928) which are set to be 100% outliers, implying that these dates will be not considered in the model iterations. Since here we used the outlier detection analysis to assess the robustness of the model, the Agreement Index is not relevant (Bronk Ramsey, 2009).

## 5. Discussion

### 5.1. Palaeoenvironment

The relationships between sea-level change and site formation processes concern aspects related to the chronology of the Middle Palaeolithic human presence in the cave, as well as peculiarities of site use by humans. However, the sequence of depositional and erosional processes that originated the sedimentary infill of the cave is still not completely known; its interpretation is still tentative and may be subject to some change when new lithologic units are uncovered.

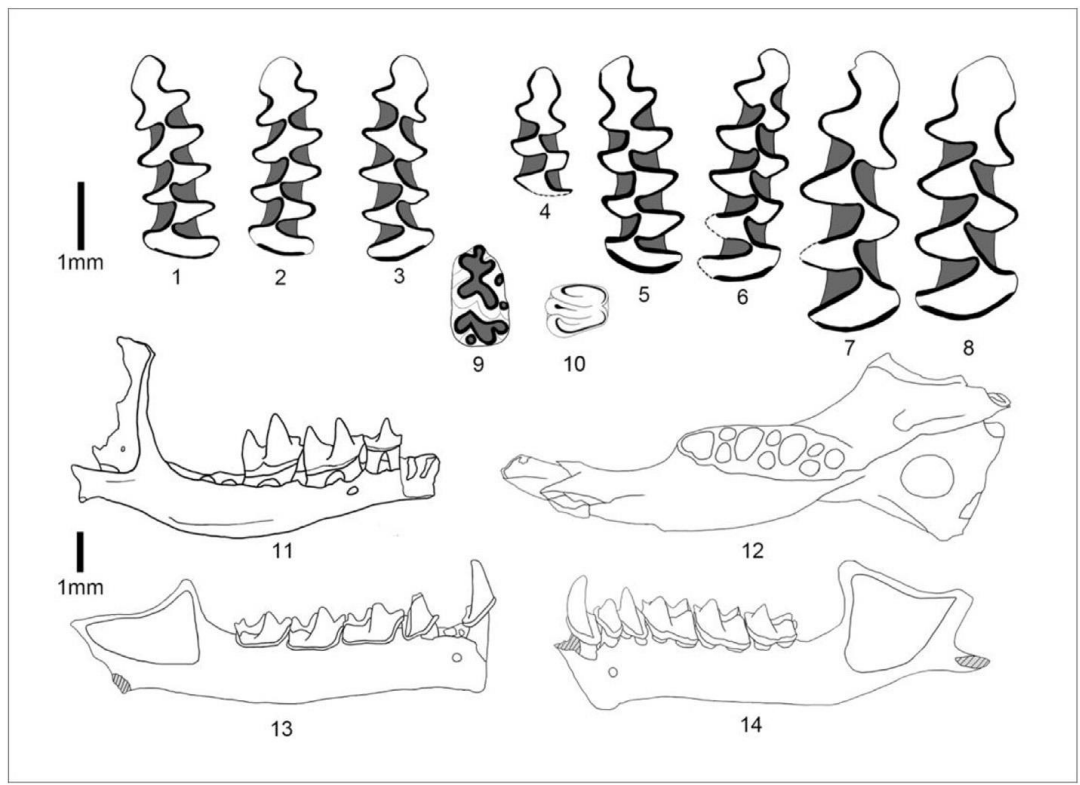

**Fig. 8.** Small mammals from Grotta dei Santi. 1e2: M. (Terricola) savii, right M1; 3: M. (Terricola) savii, left M1; 4: M. arvalis, right M1; 5: M. arvalis, left M1; 6: M. agrestis, right M1; 7e8: A. amphibius, left M1; 9: A. sylvaticus, right M1; 10: E. quercinus, left M1; 11: T. europaea, right mandible; 12: E. quercinus; right mandible; 13: R. gr. euryale-mehelyi, right mandible; 14: N. noctula, left mandible.

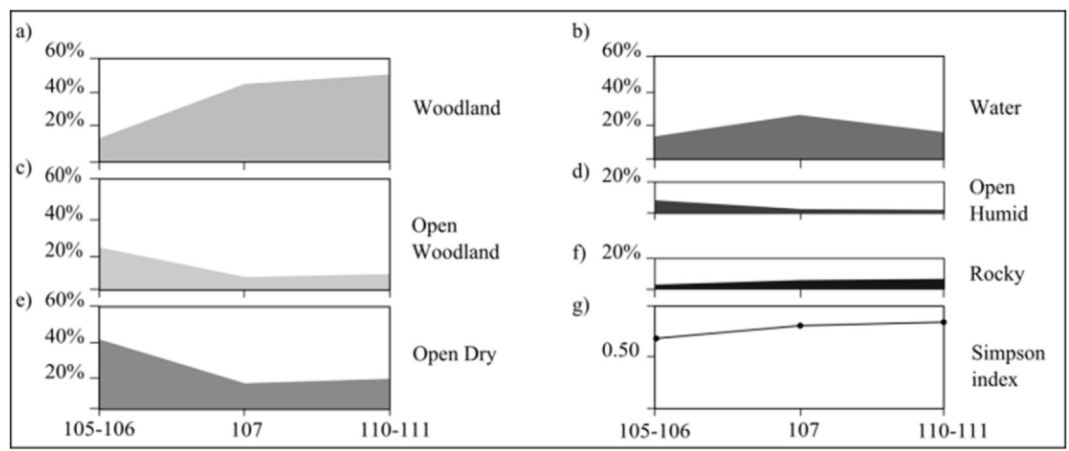

**Fig. 9**. Paleoenvironmental and palaeoclimatic reconstructions. aef: Habitat Weighting; g: Simpson index.

The accumulation of large ceiling breakdown boulders – unit 10.1, or fr following Segre (1959) - represents the first clastic deposition into the cave. It derived probably from a major phase of cliff receding, connected to some high-stand of the sea-level before theMousterian occupation. This unit thins down towards the inside of the cave suggesting that the boulders are organised in a sort of scree that originated from the outside of the cave. The stalagmite crust and bosses, if any, reportedly situated under this scree (st1 in Segre,1959, or possibly unit 10.2) would represent a previous phase of speleothem formation in a still closed cave environment.

The stratigraphic relationships between the flowstone units 10.2, 10.3 and 10.4 are still uncertain and will require accurate check during the next phases of fieldwork. At present, the only certain aspect is that the boulder scree 10.1 is covered, mostly to its inner and southern sides, by a more or less continuous flowstone (unit 10.4) that provisionally represents the base of the inner sequence.

These considerations are valid also for the patches of marine sediments (unit 10.5) adhering to the sea-facing side of the 10.1 boulders. These marine patches represent a backshore/foreshore sediments formed during a high-stand of the sea-level of at least 4e4.5m a.s.l., even if the level indicated by these sediments should be considered with care (Ferranti et al., 2006); this may appear somewhat lower than the general 7± 2m estimate suggested by Antonioli (2012) for MIS5.5. Estimates for MIS5.3 and 5.1 are too scanty to be of any utility. It can be observed that flowstone 10.4 is never covered by these sediments, which e on the contrary e overlie areas of bedrock situated at lower heights, suggesting that they were not removed from unit 10.4 by erosion. This may be evidence (though scanty) that this phase of speleothem formation occurred after the above mentioned high-stand.

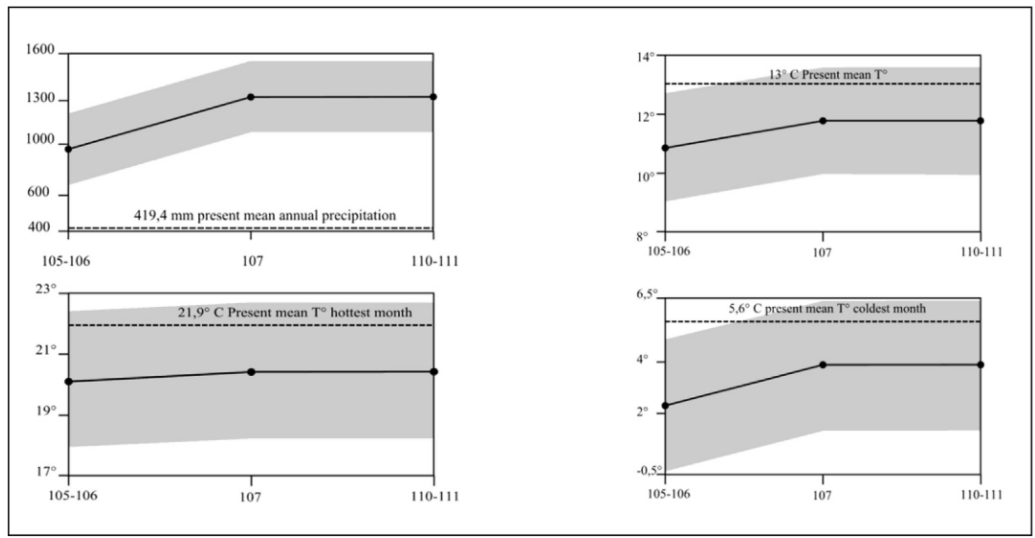

**Fig. 10.** Paleoclimatic reconstruction. Annual precipitation, Annual mean temperature, Mean temperature hottest month, Mean temperature coldest month. In hatching the CLINO (1961e1990) of Monte Argentario Weather station.

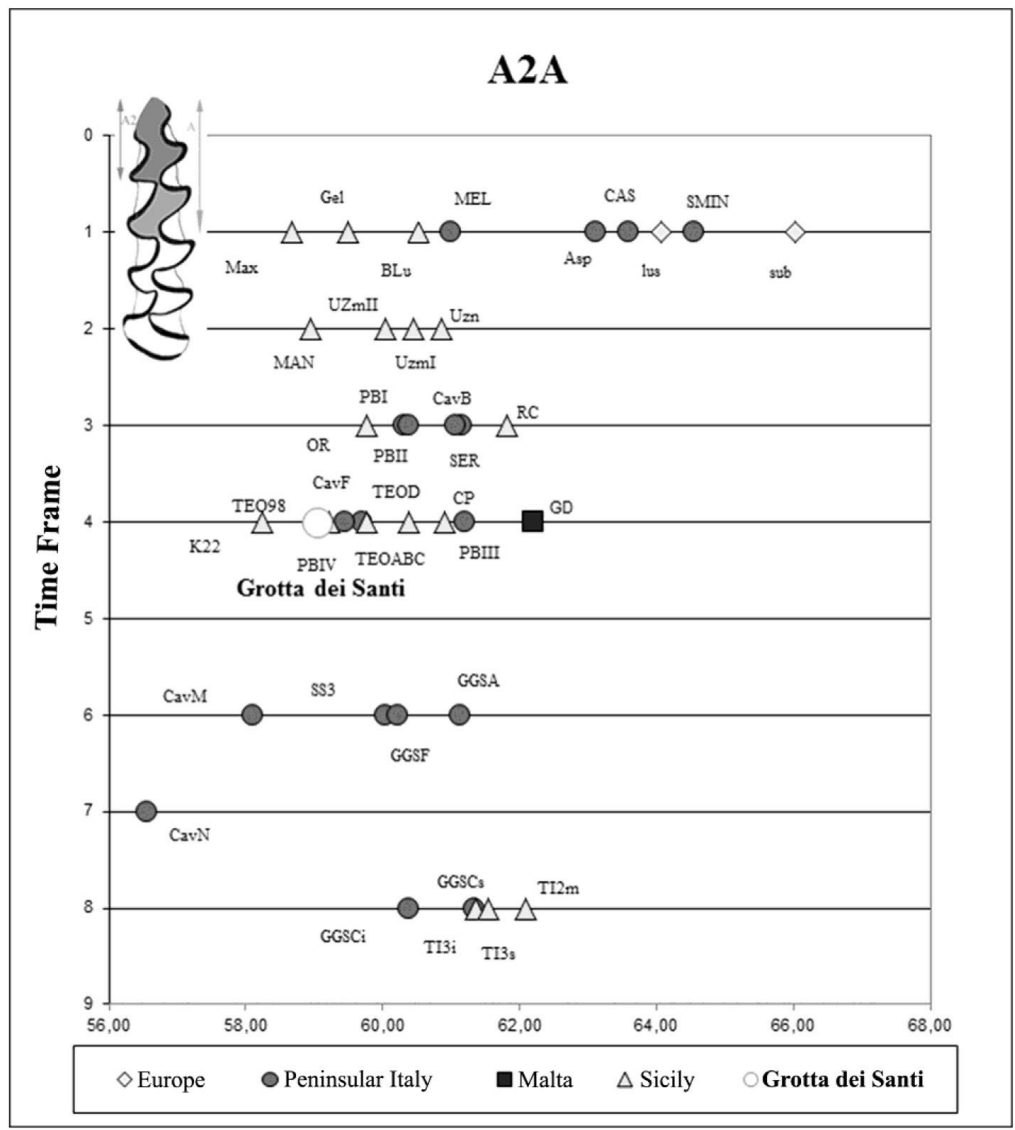

**Fig. 11.** Microtus (Terricola) savii. Comparison of A2A index of Microtus (Terricola) savii of Grotta dei Santi with other samples of Italian and European populations. Modified from Petruso et al. (2011).

| | | | | | |
|---|---|---|---|---|---|
| AL | 11 | 40.4 | 52.35 | 47.2 | 3.67 |
| BW | 11 | 2.96 | 8.12 | 4.66 | 1.74 |
| CW | 11 | 17.82 | 27.66 | 20.66 | 2.99 |
| A2A | 11 | 54.93 | 61.91 | 59.04 | 2.45 |
| DE | 11 | 29.97 | 60.34 | 42.22 | 8.32 |
| DW | 11 | 25.79 | 50.39 | 34.84 | 6.33 |
| EW | 11 | 72.5 | 90.52 | 82.94 | 5.03 |
| L | 11 | 2.3 | 2.79 | 2.51 | 0.15 |

**Table 4:** Biometric index on the first lower molar of Microtus (Terricola) savii.

Regarding the inner sequence, sea-level during the MIS5 high-stands is relevant in order to ascertain the origin of the lower-most observed unit of the inner sequence, which includes small rounded pebbles and marine mollusc shells. The origin of this unit is not yet clear: it may represent an in-cave beach deposited during a high-stand, which would fit the MIS5.1 sea-level of 7± 2m (Antonioli, 2012). Alternatively, it may have been deposited during somewhat lower stands (e.g. MIS5.3 or 5.5) by exceptionally strong southerly winter storms pushing breakers for more than 20m into the cave. Results of this process were observed a couple of times during the excavations; similar sediments accumulated on previously cleaned surfaces, to the inside of the transversal speleothem ridge of unit 10.4, which culminates at 7.2m. The dates of the overlying sediments are never older than 50,000 BP, which is consequently its youngest age, also suggesting that the unit was not deposited during the MIS5.1 high-stand, unless some still unobserved major erosion surface marks its top.

The hearth layers of unit 20.2e1004 are remarkably continuous along at least 4m. They include a complexly interlayered sequence of ash, charcoal and partly burned sediment, alternating with water- or wind-laid sterile sand layers, testifying to a cyclical and intensive occupation of the site with unvarying organisation of the living space within the cave, at least for some time.

Unit 20.3 represents several sedimentary cycles that were repeated with minor variations. Apparently, a mix of sand, silt and clay was transported into the cave by a shallow flux of medium-to low-energy water testified by the lenticular and wave ripples. Considering that the sea-shore was some kilometres far from the cave during this marine low-stand, the cyclical water and sediment input must have been due to intensive periodic rainfall. The energy of the flow decreased fast, with fining-upwards deposition of massive sand and silt, until almost pure reddish clay settled in puddling or pooling water. The reddish clay probably derived from the reworking of red soils previously formed outside the cave. Conversely, the volcanic component of the sand originated from the high-K volcanic province of southern Tuscany and northern Latium. This origin implies river transport from the volcanic areas situated to the east, to the plain area facing the cave during a low-stand of the sea-level. The frozen surface aspect of part of the grains in-dicates aeolian transport from the plain towards the entrance of the cave, also suggested by remains of laminated sandy deposits situated in shallow caves adjacent to Grotta dei Santi, 6e7m above the present-day sea-level.

Climbing dunes formed against the steep and rocky sides of the Argentario Mountain during an arid phase, likely at the end of MIS4. At present, pedogenised remains of Late Pleistocene aeolianites e not necessarily correlated with the cave ones but formed by similar processes e occur from 100 to 120m a.s.l. upwards on the side of the Argentario Mountain (Signorini, 1967). The MIS4 dunes were subsequently eroded by rainwater and transported into the cave by run-off during MIS 3, when climate shifted to more humid conditions and strong periodic rainstorms alternated with aridic periods. This shift is also indicated by the evolution of the faunal assemblage through the sequence, which starts with the occurrence of Capra ibex in unit 20.2e1004, indicating somewhat arid climate. Subsequently, the increase of cervid remains indicates a trend towards more humid conditions. The high percentage of Cervus elaphus in 20.3.2e110, associated with a small mammal assemblage dominated by Apodemus (Sylvaemus) and Eliomys quercinus, suggests a woodland environment. Bos primigenius, Equus ferus and Microtus arvalis point also to small areas of steppe or prairie, in a generically temperate climate. No large mammal data is available for the overlying units; microfaunas however suggest a moderate shift towards more open environments in units 20.3.3e107, 20.3.3e106 and 20.3.3e105.

Comparing the small mammal unit of Grotta dei Santi with the ones from the Upper Pleistocene in the Tyrrenian coastal area (Berto, 2013) we can conclude that this association can belong to the MIS3 and specifically to a shift from interstadial (20.3.3e107 and 20.3.2-110-111) to stadial (20.3.3-105-106) conditions, as highlighted by the Habitat Weighting and the Hernandez-Fernandez methods. The small mammal assemblage of Grotta dei Santi is more similar to that of the South-Tyrrenian region rather than the ones from the Central-Northen area (Berto, 2013): Monte Argentario assemblage maintains a high biodiversity, showing the presence of Woodland indicators like Eliomys quercinus and Erinaceus europaeus, even in stadial conditions.

| MPI Code | Year | Square | Unit | Comments | AMS Nr | $^{14}C$ age | 1s Err |
|---|---|---|---|---|---|---|---|
| R-EVA 927 | 2013 | H2 | 20.2–1004B (spit 1) | From the hearth | MAMS-20857 | 44,160 | 370 |
| R-EVA 1533 | 2015 | H2IV | 20.2–1004A | From the hearth | MAMS-26362 | 46,350 | 470 |
| R-EVA1534 | 2015 | H2IV | 20.2–1004A | From the hearth | MAMS-26361 | 44,400 | 390 |
| R-EVA 926 | 2013 | H2II-III | 20.2–1004A (spit 1) | Flotation in sea water | MAMS-20856 | 49,020 | 840 |
| R-EVA 928 | 2010 | F7 | 20.3.2–110 (spit 6) | Flotation in sea water | MAMS-20858 | 45,610 | 530 |
| R-EVA 1534 | 2015 | C9II | 20.3.2–110 (spit 2) | Intense dripping | MAMS-26363 | 22,840 | 70 |
| R-EVA1530 | 2015 | D9II | 20.3.2–110 (spit 1) | Intense dripping | MAMS 26360 | 33,770 | 310 |

**Table 5**: 14C results of charcoal samples from Grotta dei Santi. Dates preatreated at Mannheim using the ABOX.

| Squares | Units | Sample code | Sample ID | U (ppm) | Th (ppm) | (K) % | DR total (mGy/a) | De CAM (Gy) | Age ka |
|---|---|---|---|---|---|---|---|---|---|
| M/N 4 | 20.3.3–11 | OSL 1 | L-Eva 1452 | 2.6 ± 0.3 | 8.8 ± 0.6 | 2.0 ± 0.2 | 2.8 ± 0.2 | 100.2 ± 5.8 | 35.6 ± 3.3 |
| M/N4 | 20.3.3–24 | OSL 2 | L-Eva 1453 | 3.0 ± 0.4 | 11.0 ± 0.7 | 2.1 ± 0.2 | 3.2 ± 0.2 | 114.3 ± 8.2 | 35.4 ± 3.4 |
| H 7/8 | 20.3.2–110 | OSL 3 | L-Eva 1454 | 2.8 ± 0.5 | 10.0 ± 0.7 | 2.2 ± 0.2 | 2.9 ± 0.2 | 123.0 ± 9.7 | 42.5 ± 4.4 |
| G 7/8 | 20.3.1–125 | OSL 4 | L-Eva 1455 | 1.9 ± 0.4 | 6.9 ± 0.5 | 1.9 ± 0.1 | 2.3 ± 0.2 | 90.2 ± 6.7 | 39.7 ± 4.3 |
| H 4/5 | 20.3.1–500 | OSL 5 | L-Eva 1456 | 2.4 ± 0.5 | 7.1 ± 0.5 | 1.6 ± 0.2 | 2.4 ± 0.2 | 116.7 ± 6.7 | 49.3 ± 4.9 |
| H/I 4 | 20.3.1–900 | OSL 6 | L-Eva 1457 | 3.5 ± 0.6 | 11.4 ± 0.8 | 2.2 ± 0.2 | 2.5 ± 0.2 | 99.7 ± 8.5 | 39.2 ± 4.4 |
| E2 | 20.2–1004 | OSL 8 | L-Eva 1459 | 2.4 ± 0.4 | 9.2 ± 0.6 | 2.3 ± 0.2 | 2.6 ± 0.2 | 101.0 ± 10.0 | 38.9 ± 4.8 |

**Table 6:** Results of luminescence dating of sediment samples from Grotta dei Santi (Fig. 4).

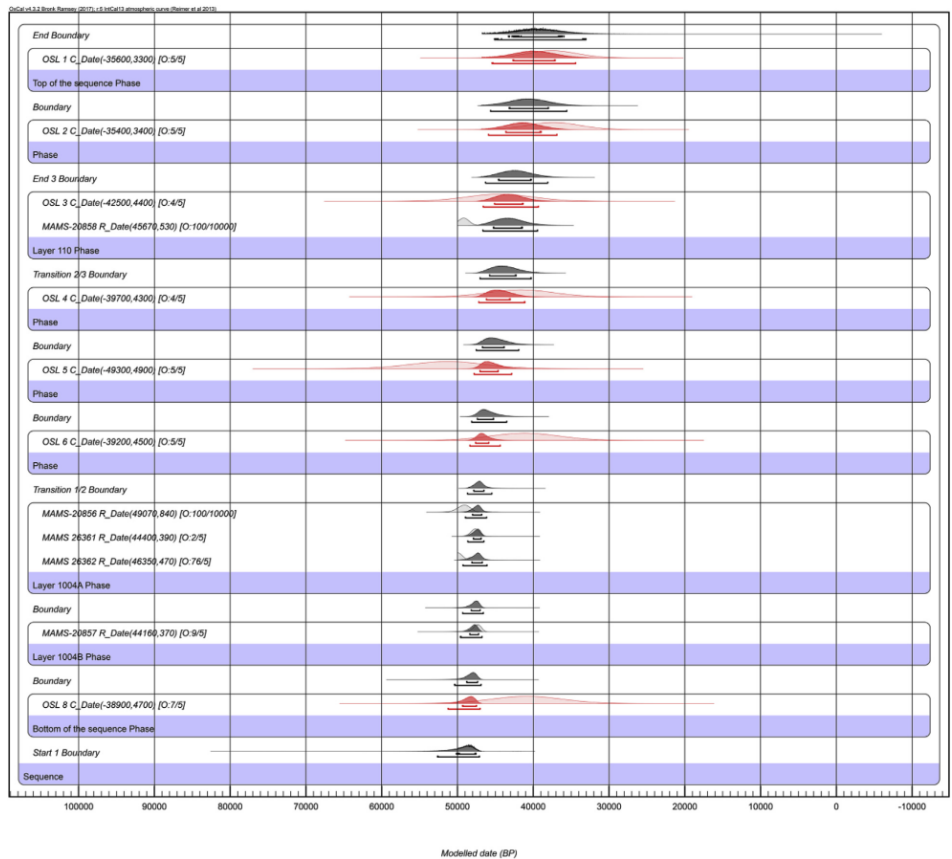

**Fig. 12.** Calibrated ages and boundaries. Bayesian model incorporating both 14C and OSL dates. The model was produced using the OxCal 4.3 program (Bronk Ramsey, 2009) and the latest International Calibration Curve, IntCal13 (Reamer et al., 2013) and each OSL age was insert as a C_Date in calendar years before 1950 using 2s error.

Concerning the presence of Nyctalus noctula concentrated in macro-unit 20.3.2-110-111, some assumptions can be made. This species is very rare in the Italian and European Upper Pleistocene panorama (Pereswiet-Soltan, 2014), in Italy it was only found at Grotta del Broion in layers R to H, (from MIS6 to the beginning of MIS3). The ecology of this species is peculiar: Nyctalus noctula usually lives in tree cavities. Its occurrence in the cave has been interpreted as the presence of a colony of this species due to the high number of individuals (MNI¼ 32) concentrated in a single macro-unit. Microtus (Terricola) savii at Grotta dei Santi represents at present the northern limit of the expansion of this species during the Upper Pleistocene. Up to now the northernmost recorded occurrence of Microtus (Terricola) savii was at the Maar of Bacciano (Kotsakis et al., 2011) in the province of Rome.

After the end of the phase testified by unit 20.3 and its faunal and anthropogenic contents, new stalagmite and flowstone crusts (unit 20.4) started growing upon the surface of unit 20.3, indicating a phase of chemical deposition in a cave environment that must have been still partially closed; it is likely that the aforementioned dunes were partially covering the present-day entrance, as also suggested by the relatively low height (less than 4m) and moderate width (about 9m) of the entrance at the level of the flowstone, which can be reconstructed following false pavement remains still adhering to the cave walls.

LGM sediments e and human/faunal presence e are not recorded at Grotta dei Santi. Although any interpretation of this lack of evidence is highly conjectural, it is not unlikely that the cave entrance was closed by the accumulation of sand dunes, which were later completely eroded by the rising sea-level during the Late Glacial and later, until recent times. Remains of flowstone 20.4 were left in place by marine erosion that acted in recent - though still unknown times - in the area of the northern and southern "corridors". Remarkable volumes of sediment of the inner sequence, mostly pertaining to unit 20.3, were removed by this phase of erosion that is still probably acting today during severe storms, whose waves enter the cave for at least 20m and 7m a.s.l.

| Sequence: Grotta dei Santi | Unmodelled (BP) | | | | Modelled (BP) | | | |
|---|---|---|---|---|---|---|---|---|
| | 68.2% Cal BP | | 95.4% Cal BP | | 68.2% Cal BP | | 95.4% Cal BP | |
| | from | to | from | to | from | to | from | to |
| **End Boundary** | | | | | 43,260 | 35,950 | 45,110 | 33,040 |
| OSL 1 C_Date(-35,600 ± 3300) | 40,850 | 34,250 | 44,140 | 30,960 | 42,670 | 37,110 | 45,430 | 34,400 |
| Top of the sequence Phase | | | | | | | | |
| **Boundary** | | | | | 43,180 | 37,990 | 45,630 | 35,570 |
| OSL 2 C_Date(-35,400 ± 3400) | 40,750 | 33,950 | 44,140 | 30,560 | 43,620 | 38,960 | 45,900 | 36,870 |
| Phase | | | | | | | | |
| **End 3 Boundary** | | | | | 44,600 | 40,260 | 46,300 | 38,080 |
| OSL 3 C_Date(-42,500 ± 4400) | 48,850 | 40,050 | 53,240 | 35,670 | 45,110 | 41,340 | 46,600 | 39,310 |
| MAMS-20858 R_Date(45,670 ± 530) | 49,760 | 48,600 | … | 48,020 | 45,260 | 41,430 | 46,650 | 39,430 |
| Layer 110 Phase | | | | | | | | |
| **Transition 2/3 Boundary** | | | | | 45,770 | 42,250 | 47,020 | 40,260 |
| OSL 4 C_Date(-39,700 ± 4300) | 45,950 | 37,350 | 50,240 | 33,070 | 46,190 | 43,030 | 47,180 | 41,100 |
| Phase | | | | | | | | |
| **Boundary** | | | | | 46,700 | 43,820 | 47,530 | 41,910 |
| OSL 5 C_Date(-49,300 ± 4900) | 56,150 | 46,350 | 61,030 | 41,470 | 47,040 | 44,620 | 47,800 | 42,830 |
| Phase | | | | | | | | |
| **Boundary** | | | | | 47,380 | 45,230 | 48,130 | 43,500 |
| OSL 6 C_Date(-39,200 ± 4500) | 45,650 | 36,650 | 50,140 | 32,170 | 47,610 | 45,870 | 48,360 | 44,370 |
| Phase | | | | | | | | |
| **Transition 1/2 Boundary** | | | | | 47,860 | 46,500 | 48,680 | 45,460 |
| MAMS-20856 R_Date(49,070 ± 840) | 49,970 | 48,260 | 50,980 | 47,530 | 48,020 | 46,790 | 48,960 | 46,160 |
| MAMS 26361 R_Date(44,400 ± 390) | 48,200 | 47,100 | 48,780 | 46,630 | 47,900 | 46,860 | 48,640 | 46,520 |
| MAMS 26362 R_Date(46,350 ± 470) | … | 49,380 | … | 48,770 | 48,090 | 46,730 | 49,270 | 46,100 |
| Layer 1004A Phase | | | | | | | | |
| **Boundary** | | | | | 48,170 | 47,010 | 49,320 | 46,590 |
| MAMS-20857 R_Date(44,160 ± 370) | 47,880 | 46,850 | 48,440 | 46,430 | 48,370 | 47,210 | 49,620 | 46,750 |
| Layer 1004B Phase | | | | | | | | |
| **Boundary** | | | | | 48,790 | 47,340 | 50,390 | 46,870 |
| OSL 8 C_Date(-38,900 ± 4700) | 45,550 | 36,150 | 50,240 | 31,470 | 49,330 | 47,500 | 51,230 | 47,010 |
| Bottom of the sequence Phase | | | | | | | | |
| **Start 1 Boundary** | | | | | 50,140 | 47,550 | 52,640 | 47,110 |

**Table 7:** Radiocarbon and OSL unmodelled and modelled dates. Calibrated ages and boundaries were calculated by using OxCal 4.3 (Bronk Ramsey, 2009) and IntCal13 (Reimer et al., 2013).

## 5.2. Chronology

Even if the quartz luminescence ages confirm the correlation of the sedimentary units and the archaeological record with MIS3, it is evident that the younger OSL ages at the bottom of the sequence have a clear problem. Hence, an age underestimation of the quartz OSL ages has to be discussed. Possible reasons might be: A) post depositional mixing and the incorporation of younger grains from the capping sterile layers into the sediments connected to the archaeological find-horizons; B) anomalous fading of the partly volcanic quartz (e.g. Tsukamoto et al., 2007). The model detected one more outlier of 76% posterior probability (MAMS-26,362, Fig. 12). In general, the start boundary ranges from 50,140 to 47,550 cal BP at 68% probability. The end boundary ranges from 43,260 to 35,950 cal BP at 68.2% probability (Table 7). The ranges in the anthropogenic unit 20.2e1004A range from 48,170 to 46,500 cal BP consistently with the strictly Mousterian characteristics of the lithic industry found in this unit. The unit 20.3.2e110 ranges from 45,770 to 40,260 cal BP. These results are also in reasonable agreement with the paleoenvironmental data; the sequence would fit reasonably within the first phases of Greenland Interstadial GI-12 spanning 46,860± 20e44,280± 60 b2k in the GICC05 tuned chronology (Blockley et al., 2012). The relatively arid conditions of unit 20.2e1004 would fit in the earliest phase of GI-12 (or in the latest phase of the previous stadial), whereas the temperate woody environment of 20.3.1e150 to 20.3.2e110, followed again by a more arid period testified by 20.3.2e110 to 20.3.3e105, would represent part of the typical D-O warm phase decline (Bond et al., 1997).

## 5.3. Human occupation of the cave

The anthropogenic units of Grotta dei Santi can be ascribed to the late Mousterian techno-cultural tradition of peninsular Italy, dated to a relatively warm period between ~50 and 40 ka ago, as suggested by geochronometric determinations supported by geological and faunal evidence.

Times and modes of the first occupation of the cave by Neandertals are still unknown, because the excavation have not yet reached the bottom of the sequence. The oldest human presence detected to date is represented by unit 20.2e1004, when several large hearths were lit during cyclical cave occupation phases over large part of the available flat surface for long time. The overlying anthropogenic units were excavated extensively and provide good evidence of the Neandertal hunting behaviour. Prime adults and mature individuals were preferred. The axial skeleton remains are underrepresented within the cave, suggesting that the first stages of butchering were carried out on the killing site, whereas the intensively fractured bones found within the cave indicate marrow extraction in the campsite.

Coprolite-rich levels systematically alternate with the anthropogenic ones in unit 20.3.2e110 and are archaeologically sterile, indicating that humans and carnivores occupied the cave cyclically in different periods. After the last human frequentation of the cave e corresponding to the top of unit 20.3.2e110, these traces of denning carnivores become dominant until the top of unit 20.3.3e105. Considering the abundance of well-preserved and intensively trampled coprolites all over the surface, it can be inferred that the cave was steadily occupied by denning hyenas for rather long periods after the Neandertals abandoned the site. Eventually, the overlying units are completely sterile of any human and animal trace, until the cave was sporadically occupied in the Roman Age.

## 5.4. Cultural and chronological framework

The Italian sites with MIS3 Mousterian occupations display a patchy distribution (Fig. S1) characterised by irregular voids and clusters that most probably do not mirror the actual Neandertal settlement distribution and demographic pattern. This distribution likely stems from various factors including geomorphology and climate constraints on Neandertal settlement choices, but also from taphonomic reasons and selective research.

Within this framework, all the sites, whose radiometric age and stratigraphic data are available in the literature, were considered in contextualising Grotta dei Santi. These sites (n. 40) represent a good corpus (for more details see Tab S1), although disparities in reliability of the chronology, in geographic distribution, continuity of human occupation and site function are evident.

The most relevant differences concern the time range of the stratigraphic sequences. Evidence of MIS 5-3 and MIS 4-3 occupation is documented in 8 and 6 sites respectively. Grotta dei Santi and 23 other ones were occupied only in MIS 3. The main exception is Grotta di San Bernardino (Colli Berici e Veneto) (Peresani et al., 2016), spanning a very broad period (MIS 7-3), although with few hiatuses.

Several sites (n¼ 15) are ascribed to MIS 3 only upon geological and/or faunal information. Radiometric dates are available for several sites (n¼ 25), but were obtained by different dating methods (TL, U/Th, OSL, ESR, 14C), and e at least in works predating year 2000e by analytic protocols that are now considered obsolete.

Interestingly, these sites are almost all caves/shelters, with the only exception of the open-air sites of Canale Mussolini (Blanc et al., 1957; Hallam and Movius, 1961), and Colle Rotondo (Villa et al., 2018) in Latium. This is a positive case, because the Mousterian levels are usually included within long sequences which, despite stratigraphic complications typical of caves, can provide better insights of the occupation times and modes than open-air sites, which are often single-layer and prone to pedogenesis and reworking.

Middle Palaeolithic evidence in Central Italy is scanty, particularly in the area facing the Adriatic Sea and in the Apennines, and generally lacks a reliable chrono-cultural framework. In the Marches, the Mousterian presence is represented by a handful of sites clustered around Ancona and chronologically limited to the "interstadials of the early last Ice Age" (Moroni et al., 2011, p.183). In situ Mousterian contexts are unknown in Umbria, despite the large quantity of surface findings, which should be broadly coeval to the Marches assemblages according to geological criteria (Moroni et al., 2011).

Several open-air localities (Galiberti, 1997) recorded in Tuscany did not always yield homogeneous and chronologically framed lithic assemblages. Some cave sequences were more or less carefully excavated and published within the 1970s; consequently, some of them were poorly documented and their lithic assemblages were studied by methods now obsolete. Only a few were investigated in more recent years (Grotta La Fabbrica, Riparo Cavanna and Grotta all'Onda).

Grotta La Fabbrica (Dini, 2011; Pitti et al., 1976; Villa et al., 2018) is a unique transitional context in Central Italy because it includes the whole sequence Mousterian e Uluzzian e Protoaurignacian. The Mousterian (layer 1) is separated from the overlying Uluzzian occupation (layer 2) by an erosive interval. According to recent studies, core reduction is based both on the Levallois and the additional concepts, whereas the bipolar technique is poorly documented. The raw material is mainly composed of pebbles from alluvial deposits (Dini, 2011; Villa et al., 2018). Several factors, including geological inferences, suggest that La Fabbrica belongs to a final phase of theMousterian. This attribution is confirmed by two OSL dates obtained for the Mousterian layer 1 (44± 2.1 ka) and the Uluzzian layer 2 (40± 1.6 ka) (Villa et al., 2018).

Further North, Riparo Cavanna e preliminarily ascribed to the first half of MIS 3 e yielded a typical Mousterian techno-complex, rich in side-scrapers and characterised by non-Levallois facies and debitage (according to Bordes's typological list) (Bachechi and Perazzi, 1996).

In North-Western Tuscany, several cave-sites (Grotta del Capriolo, Buca del Tasso, Buca della Iena, Grotta all'Onda and Grotta di Equi) are clustered on the Alpi Apuane mountain ridge (Pitti and Tozzi, 1971). These sites were excavated and studied between 1917 and 1970, yielding Levallois and denticulated late Mousterian assemblages, all attributed to the first half of MIS3. Buca della Iena and Grotta all'Onda were radiometrically dated (U/Th) to< 41,000 and< 39,300± 3200 BP respectively (Fornaca Rinaldi and Radmilli, 1968). Excavations were resumed at Grotta all'Onda in the last years, also starting a comprehensive program of radiometric dating (14C, Th/U), still in progress (Berton et al., 2003), which recently yielded, for theMousterian layers 7j3 and 7j5, two still unpublished 14C dates: 37,139± 530 (42,076e41,256 cal BP 68.2% probability, 42,462e40,724 cal BP 95.4% probability) and 36,996± 565 (42,016e41,096 cal BP 68.2% probability, 42,402e40,498 cal BP 95.4% probability).

Finally, old excavations at Grotta di Gosto (Calzoni, 1941), a cave located inland in south-eastern Tuscany, put into light a rich Mousterian collection U/Th dated to 48,000± 4000 BP (layer D, Fornaca Rinaldi and Radmilli, 1968). However, recent research at this site, including a radiometric dating program, suggests a much older chronology (Modesti, 2017).

Grotta La Fabbrica and Riparo Cavanna are the closest sites (about 34 and 68 Km as the crow flies respectively from Grotta dei Santi). Little can be said about possible comparisons between Santi and Riparo Cavanna as the assemblage from the latter is lacking a technological study and has been only generically attributed from a chronological standpoint. Some similarities between La Fabbrica and Grotta dei Santi can be detected in the technology due to the exploitation of pebbles and the use both of the Levallois and the additional methods. According to radiometric determinations, La Fabbrica was frequented by Mousterians contemporarily to the more recent occupation phase at Grotta dei Santi.

The very small size of the pebbles used in blank production and the peculiar location of Grotta dei Santi suggest similarities with the so-called Pontinian techno-complexes, especially the final Mousterian assemblages found in cave sites (e.g. Grotta Breuil) on Mt. Circeo, another promontory located southeast of Rome (Bietti and Grimaldi, 1990e91, 1996; Kuhn, 1995). Verifying this hypothesis in terms of resource procurement, production schemes, final products, etc. is one of the goals of this project.

## 6. Conclusions

Grotta dei Santi was intensively occupied by Neandertals during the MIS3 in a time span which probably covers about 10 ka, according to the first chronometric dating presented in this paper.

Geoarchaeological investigations coupled with micro- and macro fauna studies provide good evidence that the human occupation of Grotta dei Santi took place in the first half of MIS 3, during Greenland Interstadial GI-12 (Blockley, 2012).

Grotta dei Santi is providing a high-resolution picture of Neandertal behaviour in the last stages of their occupation of Central Italy. The archaeological units consist of very thin and well-separated occupation living floors constrained within a relatively short time span, each one encompassing a well identifiable occupation episode. Unlike in the so-called palimpsests, the living floors (i.e. traces of ancient camp-sites) and the activities carried out by their occupants can be reliably reconstructed, as made for other Mousterian contests (Spagnolo et al., 2016, in press; Spagnolo, 2017).

The reasons why no human occupation has been detected above unit 20.3.2e110 are still a matter of investigation. What is odd is that during the deposition of the more than 2m of sterile sediment overlying the anthropogenic units the cave seems to have been completely deserted by humans. There exists the actual possibility that from a certain time onward sediment piled up in front of the cave obstructing the entrance and hindering the access both to humans and animals. This hypothesis is corroborated by the OSL dates obtained for the upper part of the deposit, attesting to a rather abrupt interruption of sedimentation within the cave shortly after the last evidence of Neandertal occupation (units 20.3.2e110); and to a quick accumulation of the more than 2m sterile deposit covering the anthropogenic layers. Interestingly the time interval encompassed by the sterile units (modelled OSL 1 and 2 dates 42,670e37,110 and 43,620e38,960 e see Table 7) corresponds to the Uluzzian frequentation testified at La Fabbrica (OSL 40± 1.6 ka). It is possible that during the same period also Grotta dei Santi was occupied by Uluzzian and/or Protoaurignacian groups who lived nearer to the entrance of the cave (instead of in the rear part) in the deposit lost due to sea erosion. A similar arrangement of the space is known, for instance, at Grotta della Cala in Campania, where the early Upper Palaeolithic (Uluzzian and Protoaurignacian) is confined to the atrium of the cave, unlike the Mousterian which is also found in the internal series (Martini et al., 2018). The lack of human frequentation could be also interpreted in the light of a different scenario (perhaps more intriguing) related to the complex demographic dynamics occurring during the Middle to Upper Palaeolithic transition. In particular a demographic decrease of the autochthonous Neandertal populations could have caused the partial abandonment of the coastal area, leaving the territories surrounding the cave almost empty.

## Authors' contributions

A.M. and G.B. conceived and wrote the article; J.C. and F.B. macrofauna; G.M.-C. and C.B. microfauna; D.A., S.A., G.M. and A.M. lithic industry and chrono-cultural framework; T.L. and S.T. radiometric chronology; V.S. is presently the fieldwork director and together with G.B., G.C., G.M., A.M. and S.S. carried out the research fieldwork at Grotta dei Santi over the years and collaborated in writing the article; A.A., F.P., S.R., J.J.H. and S.B. collaborated in various ways in the research at Grotta dei Santi; A.G.S. carried out initial geological investigations in the cave. M.B. provided recent radiometric chronology for Grotta all'Onda.

## Acknowledgments

We aknowledge the Soprintendenza Archeologia, Belle Arti e paesaggio delle province di Siena, Arezzo e Grosseto. We are grateful to Corpo dei Vigili del Fuoco di Grosseto, Ufficio Circondariale Marittimo di Porto Santo Stefano, Ufficio Locale Marittimo di Porto Ercole, Accademia Mare Ambiente di Porto Santo Stefano, Argentario Divers di Porto Ercole, Croce Rossa di Porto Ercole, Associazione La Venta and AOG Industries di Luciano Chelli for logistic support. We are grateful to Comune di Monte Argentario, Banca di Credito Cooperativo di Castagneto Carducci, Rotary Club di Orbetello, Rotary Club di Monte Argentario and UniCoop Tirreno di Orbetello for financial support. This project has received funding from the European Research Council (ERC) under the European Union's Horizon 2020 research and innovation programme (grant agreement No. 724046); http://www.erc-success.eu/. Special thanks go to Enzo Bernabini, Andrea Bardi, Roberto Trapassi and all the firemen who took part in the fieldwork at Grotta dei Santi and to the commanders Ennio Aquilino, Mauro Caciolai and Massimo Nazareno Bonfatti, for their commitment and effort in facilitating the work and the stay in the cave of the research team. We are also grateful to people - students, researchers and enthusiasts who participated in various ways in the fieldwork at Grotta dei Santi and contributed to the achieved results. A special acknowledgement is dedicated to Stefania Campetti for resuming research at Grotta all'Onda and providing new chrono-cultural data on this important cave. Finally, we are indebted to Max Planck Society for strongly supporting this work. We are grateful to the anonymous reviewers for useful advice.

## Appendix A. Supplementary data

Supplementary data to this article can be found online at
https://doi.org/10.1016/j.quascirev.2018.11.021.

# References

Andrews, P., 2006. Taphonomic effects of faunal impoverishment and faunal mixing. Palaeogeogr. Palaeoclimatol. Palaeoecol. 241, 572e589. https://doi.org/10.1016/j.palaeo.2006.04.012.

Antonioli, F., 2012. Sea level change in western-central Mediterranean since 300 kyr: comparing global sea level curves with observed data. Alpine and Mediterranean Quaternary 25 (1), 15e23.

Arrighi, S., Bazzanella, M., Boschin, F., Wierer, U., 2016. How to make and use a bone "spatula". An experimental program based on the Mesolithic osseous assemblage of Galgenbühel/Dos de la Forca (Salurn/Salorno, BZ, Italy). Quat. Int. 423, 143e165. https://doi.org/10.1016/j.quaint.2015.11.114.

Bachechi, L., Perazzi, P., 1996. Il riparo Cavanna a Castel di Pietra (Gavorrano, Grosseto): un nuovo giacimento musteriano in toscana. In: Proceedings of the XIII Congress U.I.S.P.P., Forlì e Italia 8-14 September 1996, pp. 239e246.

Bello, S.M., Soligo, C., 2008. A new method for the quantitative analysis of cutmark micromorphology. J. Archaeol. Sci. 35, 1542e1552. https://doi.org/10.1016/j.jas.2007.10.018.

Benazzi, S., Douka, K., Fornai, C., Bauer, C.C., Kullmer, O., Svoboda, J.J., Paoa, I., Mallegni, F., Bayle, P., Coquerelle, M., Condemi, S., Ronchitelli, A., Harvati, K.,Weber, G.W., 2011. Early dispersal of modern humans in Europe and implications for Neanderthal behaviour. Nature 479, 525e528.

Benazzi, S., Slon, V., Talamo, S., Negrino, F., Peresani, M., Bailey, S.E., Sawyer, S., Panetta, D., Vicino, G., Starnini, E., Mannino, M.A., Salvadori, P.A., Meyer, M.,Paabo, S., Hublin, J.-J., 2015. The makers of the Protoaurignacian and implications for Neandertal extinction. Science 348, 793e796. https://doi.org/10.1126/science.aaa2773.

Berto, C., 2013. Distribuzione ed evoluzione delle associazioni a piccoli mammiferi nella penisola italiana durante il Pleistocene Superiore. PhD Thesis. Universita di Ferrara.

Berto, C., Boscato, P., Boschin, F., Luzi, E., Ronchitelli, A., 2017. Paleoenvironmental and paleoclimatic context during the upper palaeolithic (late upper Pleistocene) in the Italian peninsula. The small mammal record from grotta paglicci (Rignano Garganico, Foggia, southern Italy). Quat. Sci. Rev. 168, 30e41. https://doi.org/10.1016/j.quascirev.2017.05.004.

Berto, C., Luzi, E., Montanari Canini, G., Guerrreschi, F., Fontana, F., 2018. Climate and landscape in Italy during late Epigravettian. The late glacial small mammal sequence of riparo Tagliente (Stallavena di Grezzana, Verona, Italy). Quat. Sci.Rev. 184, 132e142. https://doi.org/10.1016/j.quascirev.2017.07.022.

Berton, A., Bonato, M., Borsato, A., Campetti, S., Fabbri, P.F., Mallegni, F., Perrini, L., Piccini, L., 2003. Nuove datazioni radiometriche con il metodo U/Th sulle formazioni stalagmitiche di Grotta all'Onda. Riv. Sci. Preist. LIII, 241e246.

Bietti, A., Grimaldi, S., 1990-91. Patterns of reduction sequences at Grotta Breuil: statistical analyses and comparisons of archaeological vs experimental data. Quat. Nova 1, 379e406.

Bietti, A., Grimaldi, S., 1996. Small flint pebbles and Mousterian reduction chains: the case of southern Latium (Italy). Quat. Nova 6, 237e260. Blanc, A.C., Follieri, M., Vries de, H., 1957. A first C14 date for the Würm I chronology of the Italian Coast. Quaternaria 4, 83e93

Blockley, S.P.E., Lane, C.S., Hardiman, M., Olander Rasmussen, S., Seierstad, I.K., Steffensen, J.P., Svensson, A., Lotter, A.F., Turney, C.S., Bronk Ramsey, C., 2012. Synchronization of palaeoenvironmental records over the last 60,000 years, and an extended INTIMATE event stratigraphy to 48,000 b2k. Quat. Sci. Rev. 36, 2e10. https://doi.org/10.1016/j.quascirev.2011.09.017.

Boeda, E., 2013. Techno-logique & technologie: une paleo-histoire des objets lithiques tranchants. @rcheo-editions.com, Prigonrieux, p. 259.

Bond, G., Showers, W., Cheseby, M., Lotti, R., Almasi, P., deMenocal, P., Priore, P., Cullen, H., Hajdas, I., Bonani, G., 1997. A pervasive millenial-scale cycle in North Atlantic Holocene and glacial climates. Science 278, 1257e1266. https://doi.org/ 10.1126/science.278.5341.1257.

Borzatti Von Lowenstern, 1965. La grotta-riparo di Uluzzo C (campagna di scavi 1964). Riv. Sci. Preist. 20 (1), 1e31.

Boschin, F., Crezzini, J., 2012. Morphometrical analysis on cut marks using a 3D digital microscope. Int. J. Osteoarchaeol. 22, 549e562. https://doi.org/10.1002/oa.1272.

Bronk Ramsey, C., 2009. Bayesian analysis of radiocarbon dates. Radiocarbon 51 (1),337e360. https://doi.org/10.1017/S0033822200033865.

Calzoni, U., 1941. La grotta di Gosto sulla Montagna di Cetona. Storia della scoperta e dello scavo. Studi Etruschi 25, 243e266.

Paleopedology manual. In: Catt, J.A. (Ed.), Quat. Int. 6, 1e95. Chaline, J., 1972. Les rongeurs du Pleistocene moyen et superieur de France. Cahiers de Paleontologie, Editions du Centre National de la Recherche Scientifique, pp. 1e141.

Crezzini, J., Moroni, A., 2012. Archeozoologia. La ricostruzione del comportamento umano dall'esame dei resti faunistici recuperati nei siti archeologici. Etruria Natura 9, 36e43.

Crezzini, J., Boschin, F., Boscato, P., Wierer, U., 2014. Wild cats and cut marks: exploitation of Felis silvestris in the Mesolithic of Galgenbühel/Dos de la Forca (south Tyrol, Italy). Quat. Int. 330, 52e60. https://doi.org/10.1016/j.quaint.2013. 12.056.

Dini, M., 2011. L'industria musteriana di Grotta La Fabbrica (Grosseto). Origini 23, 7e19 (Nuova serie V).

Douka, K., Grimaldi, S., Boschian, G., Del Lucchese, A., Higham, T.F.G., 2012. A new chronostratigraphic framework for the upper palaeolithic of Riparo Mochi (Italy). J. Hum. Evol. 62, 286e299. https://doi.org/10.1016/j.jhevol.2011.11.009.

Douka, K., Higham, T.F.G., Wood, R., Boscato, P., Gambassini, P., Karkanas, P., Peresani, M., Ronchitelli, A., 2014. On the chronology of the Uluzzian. J. Hum. Evol. 68, 1e13. https://doi.org/10.1016/j.jhevol.2013.12.007.

Duches, R., Nannini, N., Romandini, M., Boschin, F., Crezzini, J., Peresani, M., 2016. Identification of Late Epigravettian hunting injuries: Descriptive and 3D analysis of experimental projectile impact marks on bone. J. Archaeol. Sci. 66,88e102. https://doi.org/10.1016/j.jas.2016.01.005.

Dupuis, I., 1986. Les chiropteres du Quaternaire en France. Memoire de maîtrise en prehistoire, pp. 107e120.

Evans, E.M.N., Van Couvering, J.A.H., Andrews, P., 1981. Palaeoecology of Miocene sites in western

Kenya. J. Hum. Evol. 10, 99e116. https://doi.org/10.1016/S0047-2484(81)80027-9.

Felten, H., Helfritch, A., Storc, G., 1973. Die bestimmung der europaischen Fledermause nach der distalen Epiphyse des Humeros. Senckenbergiana biol 54,291e297.

Ferranti, L., Antonioli, F., Mauz, B., Amorosi, A., Dai Pra, G., Mastronuzzi, G., Monaco, C., Orrù, P., Pappalardo, M., Radtke, U., Renda, P., Romano, P., Sanso, P., Verrubbi, V., 2006. Markers of the last interglacial sea-level high stand along the coast of Italy: Tectonic implications. Quat. Int. 145e146, 30e54. https://doi.org/10.1016/j.quaint.2005.07.009.

Fitzsimmons, K.E., Stern, N., Murray-Wallace, C.V., 2014. Depositional history and archaeology of the central Lake Mungo lunette, Willandra Lakes, southeast Australia. J. Archaeol. Sci. 41, 349e364. https://doi.org/10.1016/j.jas.2013.08.004.

Fornaca Rinaldi, G., Radmilli, A.M., 1968. Datazione con il metodo $Th^{230}/U^{238}$ di stalagmiti contenute in depositi musteriani. Atti della Societa Toscana di Scienze Naturali serie A 75 (2), 639e646.

Freguglia, M., Gambogi, P., Milani, L., Moroni Lanfredini, A., Ricci, S., 2007. Monte Argentario (GR). Cala dei Santi: grotta dei Santi. Notiziario della Soprintendenza per i Beni Archeologici della Toscana 3, 488e491.

Fu, Q.,2, Li, H., Moorjani, P., Jay, F., Slepchenko, S.M., Bondarev, A.A., Johnson, P.L.F., Aximu-Petri, A., Prüfer, K., de Filippo, C., Meyer, M., Zwyns, N., Salazar-García, D.C., Kuzmin, Y.V., Keates, S.G., Kosintsev, P.A., Razhev, D.I.,Richards, M.P., Peristov, N.V., Lachmann, M., Douka, K., Higham, T.F.G.,Slatkin, M., Hublin, J.-J., Reich, D., Kelso, J., Viola, T.B., P€a€abo, S., 2014. Genome sequence of a 45,000-year-old modern human from western Siberia. Nature 514, 445e450. https://doi.org/10.1038/nature13810.

Galbraith, R.F., Roberts, R.G., Laslett, G.M., Yoshida, H., Olley, J.M., 1999. Optical dating of single and multiple grains of quartz from Jinmium Rock Shelter, northern Australia: part 1, experimental design and statistical models. Archaeometry 41, 339e364. https://doi.org/10.1111/j.1475-4754.1999.tb00987.x.

Galiberti, A. (Ed.), 1997. il Paleolitico e il Mesolitico della Toscana. Lalli Editore, Poggibonsi.

Grant, A., 1982. The use of tooth wear as a guide to the age of domestic ungulates. In: Wilson, B., Grigson, S., Payne, S. (Eds.), Ageing and Sexing Animal Bones from Archaeological Sites. Oxford, B. A.R., British Series, pp. 91e108.

Grimaldi, S., Santaniello, F., 2014. New insights into Final Mousterian lithic production in western Italy. Quat. Int. 350, 116e129. https://doi.org/10.1016/j.quaint.2014.03.057.

Guerin, G., Mercier, N., Adamiec, G., 2011. Dose-rate conversion factors: update.Ancient TL 29, 5e8.

Hallam, L., Movius Jr., 1961. More on upper palaeolithic archaeology. Cur.Anthrop. 2(5), 427e454.

Hammer, Ø., Harper, D.A.T., Ryan, D.P., 2001. Past: paleontological Statistics software package for education and data analysis. Paleontologia Electronica 4 (1), 1e9.

Harper, D.A.T., Hammer, O., 2006. Paleontological Data Analysis. Blackwell Publishing, Malden, MA, USA, pp. 1e351.

Hernandez Fernandez, M., 2001. Bioclimatic discriminant capacity of terrestrial mammal faunas. Global Ecol. Biogeogr. 10, 189e204. https://doi.org/10.1046/j.1466-822x.2001.00218.x.

Hernandez Fernandez, M., 2005. Analisis paleoclimatico y paleoecologico de las sucesiones de mamíferos del Plio-Pleistoceno de la Península Iberica. Servicio de Publicaciones de la Universidad Complutense de Madrid, Madrid.

Hershkovitz, I., Marder, O., Ayalon, A., Bar-Matthews, M., Yasur, G., Boaretto, E., Caracuta, V., Alex, B., Frumkin, A., Goder-Goldberger, M., Gunz, P., Holloway, R.L., Latimer, B., 12, Lavi, R., Matthews, A., Slon, V., Bar-Yosef Mayer, D., Berna, F., Bar-Oz, G., Yeshurun, R., May, H., Hans, M.G., Weber, G.W., Barzilai, O., 2015. Levantine cranium from Manot Cave (Israel) foreshadows the first European modern humans. Nature 520, 216e219. https://doi.org/10.1038/nature14134.

Higham, T.F.G., Douka, K., Wood, R., Bronk Ramsey, C., Brock, F., Basell, L., Camps, M., Arrizabalaga, A., Baena, J., Barroso-Ruíz, C., Bergman, C., Boitard, C., Boscato, P., Caparros, M., Conard, N.G., Draily, C., Froment, A., Galvan, B., Gambassini, P., Garcia-Moreno, A., Grimaldi, S., Haesaerts, P., Holt, B., Iriarte-Chiapusso, M.-J., Jelinek, A., Jorda Pardo, J.F., Maíllo-Fernandez, J.-M., Marom, A., Maroto, J., Menendez, M., Metz, L., Morin, E., Moroni, A., Negrino, F., Panagopoulou, E., Peresani, M., Pirson, S., de la Rasilla, M., Riel-Salvatore, J., Ronchitelli, A., Santamaria, D., Semal, P., 2014. The timing and spatio-temporal patterning of Neanderthal disappearance. Nature 512, 306e309. https://doi.org/10.1038/nature13621.

Hublin, J.-J., 2015. The modern human colonization of western Eurasia: when and where? Quat. Sci. Rev. 118, 194e210. https://doi.org/10.1016/j.quascirev.2014.08.011.

Hublin, J.-J., Talamo, S., Julien, M., David, F., Connet, N., Bodu, P., Vandermeersch, B., Richards, M.P., 2012. Radiocarbon dates from the Grotte du Renne and Saint-cesaire support a neandertal origin for the châtelperronian. Proc. Natl. Acad. Sci.Unit. States Am. 109, 18743e18748. https://doi.org/10.1073/pnas.1212924109.

Johnson, E., 1985. Current developments in bone technology. In: Schiffer, M.B. (Ed.), Advances in Archaeological Method and Theory. Orlando Academic Press, Orlando, pp. 157e235.

Kotsakis, T., Marcolini, F., De Rita, D., Esu, D., 2011. Three late Pleistocene small mammal faunas from the baccano maar (Rome, central Italy). Boll. Soc. Paleontol. Ital. 50, 103e110.

Kromer, B., Lindauer, S., Synal, H.-A., Wacker, L., 2013. MAMS e a new AMS facility at the curt-Engelhorn-centre for Achaeometry, Mannheim, Germany. Nucl. Instrum. Methods Phys. Res. Sect. B Beam Interact. Mater. Atoms 294, 11e13.

Kuhn, S.L., 1995. Mousterian Lithic Technology: an Ecological Perspective. Princeton University Press.

Lopez-García, J.M., Berto, C., Colamussi, V., Dalla Valle, C., Lo Vetro, D., Luzi, E., Malavasi, G., Martini, F., Sala, B., 2014. Palaeoenvironmental and palaeoclimatic reconstruction of the latest Pleistocene e Holocene sequence from Grotta del Romito (Calabria, southern Italy) using the small-mammal assemblages. Paleogeography, Paleoclimatology, Paleoecology 409, 169e179. https://doi.org/10.1016/j.palaeo.2014.05.017.

Marciani, G., Capecchi, G., Spagnolo, V., Mello Araujo, A.G., de, M., Chaves, S.A., Parenti, F., Moroni, A., 2018. Gestao, pesquisa e valorizaçao do sítio arqueologico Grotta Dei Santi (Toscana - italia). Rev. Memorare 5 (2), 86e111.

Martini, I., Ronchitelli, A., Arrighi, S., Capecchi, G., Ricci, S., Scaramucci, S., Spagnolo, V., Gambassini, P., Moroni, A., 2018. Cave clastic sediments as a tool for refining the study of human occupation of prehistoric sites: insights from the cave site of La Cala (Cilento, southern Italy). J. Quat. Sci. 33 (5), 586e596. https://doi.org/10.1002/jqs.3038.

Masini, F., Bonfiglio, L., Isacco, G., Marra, C., 1997. Large and small mammals, amphibians and reptiles from a new late Pleistocene fissure filling deposit of the Hyblean Plateau (South Eastern Sicily). Boll. Soc. Paleontol. Ital. 36, 97e122.

Merciai, G., 1910. Mutamenti avvenuti nella configurazione del litorale tra Pisa e Orbetello dal Pliocene in poi. Nistri, Pisa.

Modesti, V., 2017. Tra Toscana, Umbria e Lazio: culture e strategie di sussistenza dei cacciatore-raccoglitori neandertaliani del Monte Cetona (SI). PhD Thesis. Universita di Roma "Tor Vergata".

Moretti, E., Arrighi, S., Boschin, F., Crezzini, J., Aureli, D., Ronchitelli, A., 2015. Using 3D microscopy to Analyze experimental cut marks on animal bones produced with different stone tools. Ethnobiology Letters 6, 267e275. https://doi.org/10.14237/ebl.6.1.2015.349.

Morlan, R.-E., 1984. Toward the definition of criteria for the recognition of artificial bone alterations. Quat. Res. 22, 160e171. https://doi.org/10.1016/0033-5894(84)90037-1.

Moroni, A., Abati, F., Baldanza, A., Coltorti, M., De Angelis, M.C., Mancini, S., Moroni, B., Pieruccini, P., 2011. L'alto e medio bacino del Tevere durante il Paleolitico medio. Ipotesi sul popolamento e la mobilita dei gruppi di cacciatori-raccoglitori neandertaliani. Bollettino di Archeologia On Line 2 (2e3), 171e201.

Moroni, A., Ronchitelli, A., Arrighi, S., Aureli, D., Bailey, S.E., Boscato, P., Boschin, P., Capecchi, G., Crezzini, J., Douka, K., Marciani, G., Panetta, D., Ranaldo, F., Ricci, S., Scaramucci, S., Spagnolo, V., Benazzi, S., Gambassini, P., 2018. Grotta del Cavallo (Apulia e Southern Italy). The Uluzzian in the mirror. J. Anthropol. Sci. 96, 1e36. https://doi.org/10.4436/JASS.96004 (in press).

Moroni, A., Parenti, F., Araujo, A., Boschian, G., Boschin, F., Capecchi, G., Crezzini, J., Hublin, J.J., Marciani, G., Spagnolo, V., Talamo, S., Gambogi, P., 2015. MonteArgentario (GR). Grotta di Cala dei Santi (Concessione di Scavo). Notiziario della Soprintendenza per i Beni Archeologici della Toscana 10, 364e366.

Moroni Lanfredini, A., Freguglia, M., Bernardini, F., Boschian, G., Cavanna, C., Crezzini, J., Gambogi, P., Longo, L., Milani, L., Parenti, F., Ricci, S., 2010. Nuove ricerche alla Grotta dei Santi (Monte Argentario, Grosseto). In: NegroniCatacchio, N. (Ed.), L'Alba dell'Etruria. Fenomeni di Continuita e trasformazione nei secoli XII-VIII a.C. Ricerche e scavi, Atti del Nono Incontro di Studi Preistoria e Protostoria in Etruria - Valentano (Vt) e Pitigliano (Gr), 12-14 Settembre 2008 I, pp. 649e662.

Murray, A.S., Wintle, A.G., 2003. The single aliquot regenerative dose protocol: potential for improvements in reliability. Radiat. Meas. 37 (4e5), 377e381.

Negrino, F., Riel Salvatore, J., 2018. From Neandertals to anatomically modern humans in Liguria (Italy): the current state of knowledge. In: Borgia, V., Cristiani, E. (Eds.), Out of Italy e Advanced Studies on the Italian Palaeolithic. Sidestone Press Academics, Leida, pp. 159e180.

Nicolucci, G., 1869. Di alcune armi ed utensili in pietra rinvenuti nelle province meridionali d'Italia. Mem Atti Reale Acc. Sc. Fis. Mat. 3 (6), 130e135.

Niethammer, J., Krapp, F., 1978. Handbuck der Saugetiere Europas. Akademische Verlagsgesellschaft e Wiesbaden, pp. 1e523.

North American Commission on Stratigraphic Nomenclature, 2005. North American stratigraphic code. AAPG Bull. 89 (11), 1547e1591.

Oxilia, G., Fiorillo, F., Boschin, F., Boaretto, E., Apicella, S.A., Matteucci, C., Panetta, C., Pistocchi, R., Guerrini, F., Margherita, C., Andretta, M., Sorrentino, R., Boschian, G., Arrighi, S., Dori, I., Mancuso, G., Crezzini, J., Riga, A., Serrangeli, M.C., Vazzana, A., Salvadori, P.A., Vadini, M., Tozzi, C., Moroni, A., Feeney, R.N.M., Willman, J.C., Moggi-Cecchi, J., Benazzi, S., 2017. The dawn of dentistry in the late Upper Palaeolithic: an early case of pathological intervention at Riparo Fredian. Am. J. Phys. Anthropol. 163, 446e461. https://doi.org/10.1002/ajpa.23216.

Pagani, L., Schiffels, S., Gurdasani, D., Danecek, P., Scally, A., Chen, Y., Xue, Y., Haber, M., Ekong, R., Oljira, Mekonnen, T.E., Luiselli, D., Bradman, N., Bekele, E., Zalloua, P., Durbin, R., Kivisild, T., Tyler-Smith, C., 2015. Tracing the Route of modern humans out of Africa by using 225 human Genome sequences from Ethiopians and Egyptians. Am. J. Hum. Genet. 96 (6), 986e991. https://doi.org/10.1016/j.ajhg.2015.04.019.

Palma di Cesnola, A., 1965. Notizie preliminari sulla terza campagna di scavi nella Grotta del Cavallo (Lecce). Riv. Sci. Preist. 20, 291e302.

Palma di Cesnola, A., 1993. Il Paleolitico Superiore in Italia. Introduzione Allo Studio. Garlatti e Razzai Editori, Firenze.

Pandolfi, L., Boscato, P., Crezzini, J., Gatta, M., Moroni, A., Rolfo, M., Tagliacozzo, A., 2017. Late Pleistocene last occurrences of the narrow nosed rhinoceros Stephanorhinus hemitoechus (Mammalia, Perissodactyla) in Italy. Riv. Ital. Paleontol. Stratigr. 123 (2), 177e190. https://doi.org/10.13130/2039-4942/8300.

Peccerillo, A., 2005. Plio-quaternary Volcanism in Italy. Petrology, Geochemistry, Geodynamics. Springer, Heidelberg.

Peresani, M., Bertola, S., Del Piano, D., Benazzi, S., Romandini, M., 2018. The Uluzzian in the north of Italy. Insights around the new evidence at riparo broion Rock-shelter. Archael. Anthropol. Sci (in press).

Peresani, M., Cristiani, E., Romandini, M., 2016. The Uluzzian technology of Grotta di Fumane and its implication for reconstructing cultural dynamics in the Middle-Upper Palaeolithic transition of Western Eurasia. J. Hum. Evol. 9, 36e56. https://doi.org/10.1016/j.jhevol.2015.10.012.

Pereswiet-Soltan, A., 2014. Chirotteri italiani del Quaternario recente. PhD Thesis. Universita degli Studi di Ferrara.

Petruso, D., Locatelli, E., Surdi, G., Valle, C.D., Masini, F., Sala, B., 2011. Phylogeny and biogeography of fossil and extant Microtus (Terricola) (Mammalia, Rodentia) of Sicily and the southern Italian peninsula based on current dental morphological data. Quat. Int. 243, 192e203. https://doi.org/10.1016/j.quaint.2011.03.013.

Pitti, C., Tozzi, C., 1971. La Grotta del Capriolo e la Buca della Iena presso Mommio (Camaiore). Sedimenti, fauna, industria litica. Riv. Sci. Preist. 26 (2), 213e258.

Pitti, C., Sorrentino, C., Tozzi, C., 1976. L'industria di tipo paleolitico superiore arcaico della Grotta La Fabbrica (Grosseto). Nota preliminare. Atti della Societa Toscana di Scienze Naturali 83, 174e201.

Posth, C., Renaud, G., Mittnik, A., Drucker, D.G., Rougier, H., Cupillard, C., Valentin, F., Thevenet, C., Furtw€angler, A.,, Wißing, C., Francken, M., Malina, M., Bolus, M., Lari, M., Gigli, E., Capecchi, G., Crevecoeur, I., Beauval, C., Flas, D., Germonpre, M., van der Plicht, J., Cottiaux, R., Gely, B., Ronchitelli, A., Wehrberger, K., Grigourescu, D., Svoboda, J., Semal, P., Caramelli, D., Bocherens, H., Harvati, K.,

Conard, N.J., Haak, W., Powell, A., Krause, J., 2016. Pleistocene Mitochondrial Genomes suggest a single major dispersal of non-Africans and a late glacial population Turnover in Europe. Curr. Biol. 26 (6), 827e833. https://doi.org/10.1016/j.cub.2016.01.037.

Reimer, P.J., Bard, E., Bayliss, A., Beck, J.W., Blackwell, P.G., Bronk Ramsey, C., Grootes, P.M., Guilderson, T.P., Haflidason, H., Hajdas, I., Hatte, C., Heaton, T.J., Hoffmann, D.L., Hogg, A.G., Hughen, K.A., Kaiser, K.F., Kromer, B., Manning, S.W., Niu, M., Reimer, R.W., Richards, D.A., Scott, E.M., Southon, J.R., Staff, R.A., Turney, C.S.M., van der Plicht, J., 2013. IntCal13 and Marine13 radiocarbon age calibration curves 0-50,000 Years cal BP. Radiocarbon 55 (4), 1869e1887.https://doi.org/10.2458/azu_js_rc.55.16947.

Richter, D., Dombrowski, H., Neumaier, S., Guibert, P., Zink, A.C., 2010. Environmental gamma dosimetry with OSL of Al2O3:C for in situ sediment measurements. Radiat. Protect. Dosim. 141 (1), 27e35. https://doi.org/10.1093/rpd/ncq146.

Romagnoli, F., Martini, F., Sarti, L., 2015. Neanderthal Use of Callista chione Shells as Raw Material for Retouched Tools in South-east Italy: analysis of Grotta del Cavallo Layer L Assemblage with a New Methodology. J. Archaeol. Method Theor 22 (4), 1007e1037. https://doi.org/10.1007/s10816-014-9215-x.

Romagnoli, F., Baena, J., Sarti, L., 2016. Neanderthal retouched shell tools and Quina economic and technical strategies: an integrated behavior. Quat. Int. 407, 29e44. https://doi.org/10.1016/j.quaint.2015.07.034.

Salvagnoli, A., Marchetti, A., 1843. Armi e utensili nella grotta de'Santi presso il Monte Argentario. In: Atti 5 Riunione Degli Scienziati Italiani, p. 264.

Segre, A.G., 1959. Giacimenti pleistocenici con fauna e industria litica a Monte Argentario (Grosseto). Riv. Sci. Preist. 14, 1e18.

Sevilla, P., 1988. Estudio paleontologico de los Quiropteros del Cuaternario espagnol. PhD thesis. Istitut paleontologic Dr. M. Crusafont.

Signorini, F., 1967. Note Illustrative Alla Carta Geologica D'Italia, Foglio 135 -Orbetello. Servizio Geologico d'Italia, Roma, p. 28.

Spagnolo, V., 2017. Studio delle strategie insediative del Paleolitico medio in Italia centro-meridionale. PhD Thesis. Universita degli Studi di Siena.

Spagnolo, V., Marciani, G., Aureli, D., Berna, F., Toniello, G., Astudillo, F.J., Boscato, P., Ronchitelli, A., 2018. Neanderthal activity and resting areas from Stratigraphic Unit 13 at the Middle Palaeolithic site of Oscurusciuto (Ginosa - Taranto, Southern Italy). Quat. Sci. Rev. 217, 169e193.

Spagnolo, V., Marciani, G., Aureli, D., Berna, F., Boscato, P., Ranaldo, F., Ronchitelli, A., 2016. Between hearths and volcanic ash: the SU 13 palimpsest of the Oscurusciuto rock shelter (Ginosa Southern Italy): analytical and interpretative questions. Quat. Int. 417, 105e121. https://doi.org/10.1016/j.quaint.2015.11.046.

Tsukamoto, S., Murray, A.S., Huot, S., Watanuki, T., Denby, P.M., Bøtter-Jensen, L., 2007. Luminescence property of volcanic quartz and the use of red isothermal TL for dating tephras. Radiat. Meas. 42 (2), 190e197. https://doi.org/10.1016/j.radmeas.2006.07.008.

Van der Meulen, A.J., 1972. Middle Pleistocene smaller mammals from the Monte peglia (Orvieto, Italy), with special reference to the phylogeny of Microtus (Arvicolidae, Rodentia). Quaternaria 17, 1e144.

Vicino, G., 1972. Gli scavi preistorici nell'area dell'Ex-Casino dei Balzi Rossi (nota preliminare). Rivista Ingauna e Intemelia 27, 77e97.

Villa, P., Mahieu, E., 1991. Breakage patterns of human long bones. J. Hum. Evol. 21, 27e48. https://doi.org/10.1016/0047-2484(91)90034-S.

Villa, P., Pollarolo, L., Conforti, J., Marra, F., Biagioni, C., Degano, I., Lucejko, J.J., Tozzi, C., Pennacchioni, M., Zanchetta, G., Nicosia, C., Martini, M., Sibilia, E.,Panzeri, L., 2018. From Neandertals to modern humans: new data on the Uluzzian. PloS One 13 (5) e0196786. https://doi.org/10.1371/journal.pone.0196786.

Wealbroeck, C., Labeyrie, L., Michel, E., Duplessy, J.C., Lambeck, K., Mcmunus, J.F.,Balbon, E., Labracherie, M., 2002. Sea-level and deep water temperature changes derived from benthic foraminifera isotopic records. Quat. Sci. Rev. 21, 295e305. https://doi.org/10.1016/S0277-3791(01)00101-9.

Zanchetta, G., Giaccio, B., Bini, M., Sarti, L., 2018. Tephrostratigraphy of Grotta del Cavallo, Southern Italy: insights on the chronology of Middle to Upper Palaeolithic transition in the Mediterranean. Quat. Sci. Rev. 182, 65e77. https://doi.org/10.1016/j.quascirev.2017.12.014.